\documentclass[twocolumn,twoside,slac_two]{revtex4}
\usepackage{graphicx}
\usepackage{fancyhdr}
\usepackage{epsfig}
\usepackage{psfrag}
\usepackage{slashed}

\newcommand{\slsh}{\!\!\!/}
\newcommand{\ba}[0]{\begin{eqnarray}}
\newcommand{\ea}[0]{\end{eqnarray}}

\newcommand{\PIQ}{\ensuremath{\Pi(Q^{2})} }

\newcommand{\MD}{\ensuremath{\mathcal{D}} }
\newcommand{\MK}{\ensuremath{\mathcal{K}} }
\newcommand{\MU}{\ensuremath{\mathcal{U}} }

\newcommand{\omP}{\omega_{\Pi}}
\newcommand{\omD}{\omega_{\MD}}

\newcommand{\QQ}{\ensuremath{Q^{2}}}

\begin{document}

\title{\bf Perturbative and non-perturbative QCD}

%

\author{C.J. Maxwell}
\affiliation{Institute for Particle Physics Phenomenology (IPPP), Durham University, U.K.}
\begin{abstract}
In these lectures we give a concise introduction to the ideas
of renormalon calculus in QED and QCD. We focus in particular on the
example of the Adler $D$ function of vacuum polarization, and
on relations between perturbative renormalon ambiguities and corresponding
non-perturbative Operator Product Expansion (OPE) ambiguities. Recent
work on infrared freezing of Euclidean observables is also discussed,

\end{abstract}

\maketitle

\thispagestyle{fancy}
\section{Introduction}
In these lectures I will discuss the large-order behaviour 
of perturbation theory. The field theory perturbation series in the
renormalised coupling is not convergent, and in $n$th order
the coefficients exhibit $n$! growth. The sources of this divergent
behaviour are instantons, which are connected with the combinatoric
growth in the number of Feynman diagrams, and renormalons which
are associated with individual diagrams containing chains of
fermion bubbles. We will focus on vacuum polarization in QED and
QCD, and will analyse the Borel plane singularities which
arise due to these chain diagrams, so-called ultraviolet (UV) and
infrared (IR) renormalons. In QED the UV renormalons render 
perturbative results ambiguous, whereas the ambiguity in perturbative
QCD arises from IR renormalons. The IR ambiguities in perturbative 
QCD are associated with non-logarithmic UV divergences in the non-perturbative
Operator Product Expansion (OPE), and can cancel between the perturbative
and non-perturbative sectors. We shall particularly focus on the deep connections
between the perturbative and non-perturbative sectors, reporting in the final
section on some recent work on infrared frezing \cite{r1} which suggests that IR and
UV renormalons conspire between them to provide finiteness and continuity
avoiding a Landau Pole in the coupling. For an excellent review on renormalon
calculus see Ref.\cite{r2}, and for a review on connections between perturbation
theory and non-perturbative power corrections see Ref.\cite{r3}. Recommended texts
on quantum field theory are Refs.\cite{r4,r5,r6}.
We begin with a brief introduction
to QCD, and introduce the vacuum polarization on which we will largely focus. 

\section{Introduction to QCD}
Quantum Chromodynamics (QCD) is a non-abelian gauge theory of interacting quarks
and gluons. The gauge group is SU$(N_c)$, and there are ${N}_{c}^{2}-1$ gluons. Experimental indications are
that ${N}_{c}=3$. The Lagrangian density is
\begin{eqnarray}
{\cal{L}}_{QCD}={\bar{\psi}}(i{\gamma}^{\mu}{\partial}_{\mu}-m){\psi}-g_s({\bar{\psi}}{\gamma}^{\mu}{T}^{a}\psi){G}^{a}_{\mu}-\frac{1}{8}{G}^{a}_{\mu\nu}
{G}^{\mu\nu}_{a}.
\end{eqnarray}
Here $a=1,2,3,\ldots,8$, and ${T}_{a}={\lambda}_{a}/2$ are the generators of SU$(3)$, the $3\times{3}$
Gell-Mann $\lambda$-matrices, satisfying
\begin{eqnarray}
[{T}_{a},{T}_{b}]=i{f}_{abc}{T}_{c}.
\end{eqnarray}
The quark fields carry colour, R, G, B, and transform as a triplet in the fundamental representation
\begin{eqnarray}
\psi(x)=\pmatrix{\psi_{R}(x)\cr{\psi}_{G}(x)\cr{\psi}_{B}(x)\cr}.
\end{eqnarray}
${\cal{L}}_{QCD}$ is invariant under local SU$(3)$ gauge transformations
\begin{eqnarray}
\psi(x)\rightarrow U(x)\psi={e}^{i{T}^{a}{\alpha}_{a}(x)}\psi(x).
\end{eqnarray}
The field strength tensor ${G}^{a}_{\mu\nu}$ contains the abelian (QED) result and an extra
term proportional to the structure constants $f^{abc}$ which are responsible for three and four-point
self-interactions of gluons, not present for photons in QED.
\begin{eqnarray} 
{G}^{a}_{\mu\nu}={\partial}_{\mu}{G}^{a}_{\nu}-{\partial}_{\nu}{G}^{a}_{\mu}+{g}_{s}{f}^{abc}{G}^{b}_{\mu}{G}^{c}_{\nu}.
\end{eqnarray}
For QCD (but not QED) one also needs to include unphysical {\it ghost} particles. These
are scalar Grassmann (anti-commuting) fields  needed to cancel unphysical polarization states for the gluons.
The required Fadeev-Popov extra term in ${\cal{L}}_{QCD}$ is
\begin{eqnarray}
{\cal{L}}_{\rm{ghost}}={\bar{\eta}}^{a}(-{\partial}^{2}{\delta}^{ac}-{g}_{s}{\partial}^{\mu}{f}^{abc}{G}^{b}_{\mu}){\eta}^{c}.
\end{eqnarray}
In both QED and QCD one needs also to include a gauge fixing term if inverse propagators are to
be defined.
\begin{eqnarray}
{\cal{L}}_{\rm{gauge-fixing}}=\frac{1}{2\xi}{({\partial}^{\mu}{G}^{a}_{\mu})}^{2}.
\end{eqnarray}
There is only one other gauge-invariant structure that we could add involving the {\it dual} field
strength tensor ${\tilde{G}}^{a}_{\mu\nu}$,
\begin{eqnarray}
{\cal{L}}_{\theta}=\frac{\theta{g}_{s}^{2}}{64{\pi}^{2}}{\tilde{G}}^{a,\mu\nu}{G}^{\rho\sigma}_{a}.
\end{eqnarray}
This is a total derivative and so produces no effects at the perturbative level.
However, if $\theta\neq{0}$ non-perturbative effects would induce a CP-violating electric dipole moment for 
the neutron, experimental constraints on this provide a bound $|\theta|<3.{10}^{-10}$.\\ 
\subsection{QED vacuum polarization}
\begin{figure*}[t]
\centering
\epsfig{file = 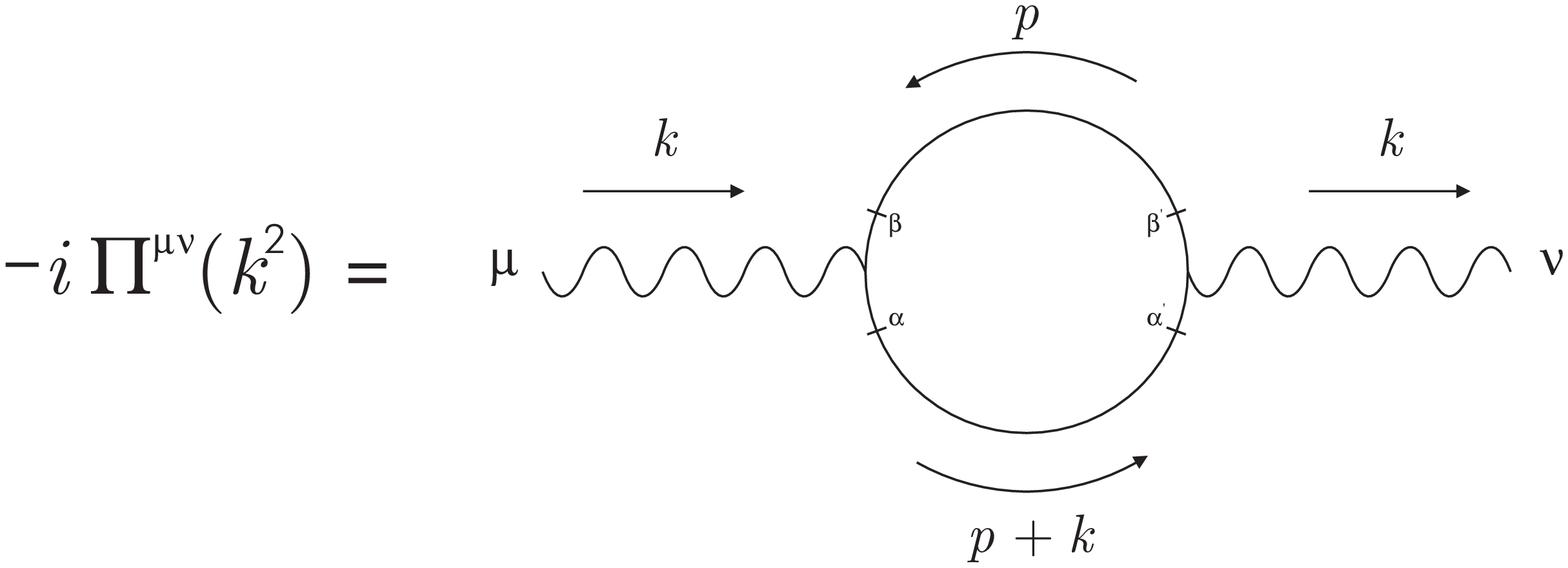, width = 9cm}
\caption{One-loop diagram for QED vacuum polarization}
\end{figure*}
A crucial ingredient in our later discussions will be the one-loop vacuum
polarization diagram shown in Fig.1.
Using the Feynman rules the diagram is
given by
\begin{eqnarray}
-i\Pi^{\mu\nu}(k^{2})&\equiv&
(-1)\int\frac{d^{4}p}{(2\pi)^{4}}\Bigg[(-ie\gamma^{\mu})_{\alpha\beta}\frac{i(p\slsh+m)_{\beta\beta'}}{p^{2}-m^{2}}
\nonumber \\
&\times&(-ie\gamma^{\nu})_{\beta'\alpha'}\frac{i(p\slsh+k\slsh+m)_{\alpha'\alpha}}{[(k+p)^{2}-m^{2}]}\Bigg],
\end{eqnarray}
Here $e$ is the QED coupling, $\alpha={e}^{2}/4\pi$. The integral is divergent and requires ``regularization''.
The most widely applied method
is ``dimensional regularization'' , in which the integral is performed in $D=4-2\varepsilon$ spacetime dimensions
and then the limit $\varepsilon\rightarrow{0}$ is taken. 
We need to evaluate
\begin{eqnarray}
-i\Pi^{\mu\nu}(k^{2})=-e^{2}\int
\frac{d^{D}p}{(2\pi)^{D}}\frac{\textrm{Tr}[\gamma^{\mu}p\slsh\gamma^{\nu}(p\slsh+k\slsh)]}{p^{2}(k+p)^{2}}.
\end{eqnarray}
The continuation to $D$-dimensions endows the dimensionless coupling $e$
with a mass dimension, $[e]=2-\frac{D}{2}=\varepsilon$, so one needs to replace $e\rightarrow{e}{\mu}^{\varepsilon}$.
One finally obtains
\begin{eqnarray}
-i{\Pi}^{\mu\nu}(k^2)&=&-i\frac{\alpha}{3\pi}[{k}^{\mu}{k}^{\nu}-{g}^{\mu\nu}{k}^{2}]\Bigg{(}-\frac{1}{\varepsilon}+
\ln\frac{-k^2}{\mu^2} 
\nonumber \\
&-&(\ln{4\pi}-{\gamma}_{E})-\frac{5}{3}\Bigg{)}.
\end{eqnarray}
Here the divergence as $\varepsilon\rightarrow{0}$ is contributed by $\Gamma(\varepsilon)=\frac{1}{\varepsilon}-{\gamma}_{E}+O(\varepsilon)$
, where ${\gamma}_{E}\approx{0.5722}$ is the Euler constant.
Counterterms are introduced to remove the $1/\varepsilon$ divergences (see next section),
the finite contribution they also cancel is arbitrary and determines the subtraction procedure. Modified
minimal subtraction (${\overline{MS}}$) absorbs the $\ln{4\pi}-{\gamma}_{E}$ term, minimal subtraction
($MS$) does not.
\subsection{Renormalization and running coupling}
One needs to distinguish between the {\it bare} Lagrangian ${\cal{L}}_{B}$ and the renormalised
Lagrangian ${\cal{L}}_{R}$. Bare and renormalized fields and couplings are related by {\it infinite} renormalization
factors $Z_i$. In massless QCD for instance one has
\begin{eqnarray}
{\psi}_{B}={Z}_{2}^{1/2}{\psi}_{R}\;\;,{G}^{\mu,a}_{B}={Z}_{3}^{1/2}{G}^{\mu,a}_{R}\;\;,{g}_{sB}={Z}_{g}{g}_{sR}.
\end{eqnarray}
This infinite reparametrization can be implemented by introducing counterterms into the Lagrangian.
These will contain counterterm coefficients, by choosing suitable values for these coefficients proportional to $1/{\varepsilon}$, 
the divergent parts present in loop calculations can be cancelled. One finds that the renormalised coupling runs logarithmically
with the renormalization scale $\mu$, and satisfies the beta-function equation
\begin{eqnarray}
\frac{da}{d\ln \mu}=\beta{(a)}=-b{a}^{2}(1+ca+{c}_{2}{a}^{2}+{c}_{3}{a}^{3}+\ldots).
\end{eqnarray}
Here $a(\mu^2)={\alpha}_{s}(\mu^2)/{\pi}={g}_{s}(\mu^2)/4{\pi}^{2}$. 
The terms up to and including $c_3$ have been computed in the ${\overline{MS}}$ renormalization
scheme. The first two coefficients are scheme-independent
\begin{eqnarray}
b=\frac{(11{N}_{c}-2{N}_{f})}{6}\;,\;c=\frac{(153-19{N}_f)}{12b}
\end{eqnarray}
The first $N_c$-dependent contribution to $b$ arises from gluon and ghost vacuum polarization contributions,
and the second $N_f$-dependent contribution is (up to a group theory factor of $T(R)=1/2$) just the QED
vacuum polarization contribution considered earlier. $N_f$ is the number of active quark flavours (fermion
species in QED). For SU$(3)$ QCD with $N_f<33/2$, $b>0$, and $a(\mu^2)\rightarrow{0}$ as ${\mu}^{2}\rightarrow{\infty}$
(Asymptotic Freedom). At the one-loop level the solution of the equation is
\begin{eqnarray}
a(\mu^2)=\frac{2}{b\ln (\mu^2/{\Lambda}^2)}.  
\end{eqnarray}
At two-loops the solution may be written in terms of the Lambert $W$-function \cite{r7} defined
implicitly by $W(z)exp(W(z))=z$ 
\begin{eqnarray}
a(\mu^2)&=&-\frac{1}{c[1+W(z(\mu)]}
\nonumber \\
z(\mu)&=&-\frac{1}{e}{\left(\frac{\mu}{\Lambda}\right)}^{-b/c}. 
\end{eqnarray}
At higher loops $a(\mu^2)$ will depend on the choices of the non-universal beta-function
coefficients $c_2,c_3,\ldots$. A special choice is an {\it `t Hooft} scheme \cite{r8} where ${c}_{2}={c}_{3}\ldots={c}_{n}=0$
in which case $a(\mu^2)$ may be written in terms of $W(z)$ as above.

\subsection{Vacuum polarization and the ${e}^{+}{e}^{-}$ hadronic total cross section}
Consider ${e}^{+}(p_1)+{e}^{-}(p_2)\rightarrow{X}$, where $X$ is a hadronic system having
total momentum $q={p}_{1}+{p}_{2}$. The squared ${e}^{+}{e}^{-}$ cm energy is $s=q^2$. The relevant amplitude is
\begin{eqnarray}
A({e}^{+}{e}^{-}\rightarrow{X})=\frac{2\pi{e}^{2}}{s}{\bar{v}}(p_1){\gamma}^{\mu}u(p_2)\langle{X}|{J}_{\mu}|0\rangle.
\end{eqnarray}
Here $J_\mu$ is the electromagnetic current for quarks
\begin{eqnarray}
J^\mu=\sum_{f}{Q}_{f}{\bar{\psi}}_{f}{\gamma}^{\mu}{\psi}_{f},
\end{eqnarray}
summed over active quark flavours. The ${e}^{+}{e}^{-}$ total hadronic cross section is
then given by
\begin{eqnarray}
\sigma&=&\sum_{X}\sigma({e}^{+}{e}^{-}\rightarrow{X})
\nonumber \\
&=&\frac{8{\pi}^{2}{\alpha}^{2}}{s^3}{{l}}_{\mu\nu}\sum_{X}{(2\pi)}^{4}{\delta}^{(4)}({p}_{1}+{p}_{2}-q)
\nonumber \\
&\times&\langle{X}|{J}^{\nu}(0)|0\rangle{\langle X|{J}^{\mu}(0)|0\rangle}^{\ast}.
\end{eqnarray}
Here the leptonic tensor ${l}_{\mu\nu}$ is given by
\begin{eqnarray}
{l}_{\mu\nu}=\frac{1}{4}{\rm{Tr}}[{\slashed{p}}_{1}{\gamma}_{\mu}{\slashed{p}}_{2}{\gamma}_{\nu}].
\end{eqnarray}
Using the {\it optical theorem} the $\sum_X$ term may be related to the imaginary part of the photon propagator.
We may write
\begin{eqnarray}
&\sum_{X}&{(2\pi)}^{4}{\delta}^{(4)}({p}_{1}+{p}_{2}-q)\langle{0}|{J}_{\mu}(0)|X\rangle{\langle{0}|{J}_{\nu}(0)|X\rangle}^{\ast}
\nonumber \\
&=& \int{{d^4}x}{e}^{iq.x}\langle{0}|[{J}_{\mu}(x),{J}_{\nu}(0)]|0\rangle
\nonumber \\
&=& \frac{1}{8{\pi}^{2}} Im{\Pi}^{\mu\nu}(s).
\end{eqnarray}
Here ${\Pi}^{\mu\nu}(Q^2)$ is the vacuum polarization function
\begin{eqnarray}
{\Pi}^{\mu\nu}(Q^2)=16{\pi}^{2}i\int{d^4x}{e}^{iq.x}\langle{0}|T({J}_{\mu}(x){J}_{\nu}(0))|0\rangle
\end{eqnarray}
Here $Q^2=-q^2>0$.
Conservation of $J_\mu$, ${\partial}_{\mu}{J}^{\mu}=0$ then  dictates the tensor structure
\begin{eqnarray}
{\Pi}^{\mu\nu}(Q^2)=({q}_{\mu}{q}_{\nu}-{g}_{\mu\nu}{q}^{2}){\Pi}(Q^2).
\end{eqnarray}
The quark-loop vacuum polarization diagram we considered earlier contains $\ln(-s/{\mu}^{2})$ which has
an Im part of $i\pi$. Combining with QCD colour factors, and taking into account the fractional charges of quarks one
finds
\begin{eqnarray}
Im[{\Pi}^{(0)\mu\nu}(s)]=\frac{4\pi}{3}3\sum_{f}{Q}_{f}^{2}.
\end{eqnarray}
One then finds for the total hadronic cross-section
\begin{eqnarray}
\sigma({e}^{+}{e}^{-}\rightarrow{\rm{hadrons}},s)=\frac{4\pi{\alpha}^{2}}{3s}3\sum_{f}{Q}_{f}^{2}.
\end{eqnarray}
A convenient observable to measure at ${e}^{+}{e}^{-}$ colliders such as LEP
is the ${R}_{{e}^{+}{e}^{-}}$ ratio, defined by
\begin{eqnarray}
{R}_{{e}^{+}{e}^{-}}(s)=\frac{\sigma({e}^{+}{e}^{-}\rightarrow{\rm{hadrons}},s)}{\sigma({e}^{+}{e}^{-}\rightarrow{\mu}^{+}{\mu}^{-},s)}.
\end{eqnarray}
The ${\mu}^{+}{\mu}^{-}$ cross-section can be directly measured in the experiment,
\begin{eqnarray}
\sigma({e}^{+}{e}^{-}\rightarrow{\mu}^{+}{\mu}^{-},s)=\frac{4\pi{\alpha}^{2}}{3s}.
\end{eqnarray}
The total hadronic cross-section differs from the ${\mu}^{+}{\mu}^{-}$ point cross-section by the
factor $3$ ($={N}_{c}$ colours for each quark/anti-quark), and $\sum_{f}{Q}_{f}^{2}$ which takes into account
the fractional quark charges. Taking the ratio one finds
\begin{eqnarray}
{R}_{{e}^{+}{e}^{-}}(s)=3\sum_{f}{Q}_{f}^{2}[1+{\cal{R}}(s)].
\end{eqnarray}
Here ${\cal{R}}(s)={\cal{R}}_{PT}(s)+{\cal{R}}_{NP}(s)$ corresponds to QCD corrections
to the parton model result. The perturbative (PT) contribution is of the form 
\begin{eqnarray}
{\cal{R}}_{PT}(s)=a(s)+{r}_{1}a^{2}(s)+{r}_{2}{a}^{3}(s)+\ldots.
\end{eqnarray}
Here $a(s)$ denotes the coupling in the $\overline{MS}$ scheme, the $r_1$ and $r_2$ corrections have
been computed in the $\overline{MS}$ scheme. The non-perturbative (NP) component arises
from the Operator Product Expansion (OPE).\\

Only ${\Pi}(Q^2)-\Pi (0)$ is observable and so 
it is convenient to take a logarithmic derivative
with respect to $Q^2$ and define the Adler $D$ function,
\begin{eqnarray}
D(Q^2)=-\frac{3}{4}Q^2\frac{d}{dQ^2}{\Pi}(Q^2).
\end{eqnarray}
This has the same parton model expression as $R_{{e}^{+}{e}^{-}}(s)$ with the perturbative
corrections ${\cal{R}}_{PT}(s)$ replaced by ${\cal{D}}_{PT}(s)$,
\begin{eqnarray}
{\cal{D}}_{PT}(Q^2)=a(Q^2)+{d}_{1}{a}^{2}(Q^2)+{d}_{2}{a}^{3}(Q^2)+\ldots.
\end{eqnarray}
${\cal{R}}(s)$ is related to ${\cal{D}}(-s)$ , by analytical continuation from Euclidean
to Minkowskian momenta. One can write the dispersion relation
\begin{eqnarray}
{\cal{R}}(s)=\frac{1}{2\pi i}\int^{-s+i\epsilon}_{-s-i\epsilon}{dt}\frac{{\cal{D}}(t)}{t}.
\end{eqnarray}
One can also write this as a circular contour integral in the complex-$s$ plane
\begin{eqnarray}
{\cal{R}}(s)=\frac{1}{2\pi}\int_{-\pi}^{\pi}{d\theta}{\cal{D}}(s{e}^{i\theta}).
\end{eqnarray}
\section{Large-order behaviour of PT- Instantons and Renormalons}
In 1952 Dyson presented an argument \cite{r9,r10} that QED perturbation series are
divergent series with coefficients growing like $n!$ in $n$th order. Consider
\begin{eqnarray}
f(e^{2})&=&\sum_{n=0}^{\infty}f_{n}e^{2n}.
\end{eqnarray} 
Assuming that $f(e^2)$ converges for ${e}^{2}>0$, implies that $f(e^2)$ is analytic
at $e^2=0$, and must converge for some negative values of $e^2$. For $e^2<0$, however, like charges
attract and unlike charges repel !  The vacuum state is then unstable and it becomes energetically
favourable to create more and more ${e}^{+}{e}^{-}$ pairs. Consider a system of $N$ interacting electrons,
the energy will be
\begin{eqnarray}
E\sim{N}T+\frac{1}{2}{N}^{2}V{e}^{2}.
\end{eqnarray}
Here $T$ is the mean kinetic energy, $V$ the mean coulomb potential and $\frac{1}{2}(N^2-N)\sim\frac{1}{2}N^2$ 
counts the interacting ${e}^{+}{e}^{-}$ pairs. For $e^2>0$ the energy (wrt $N$) is bounded from below and there
exists a stable minimum, the vacuum state, at $N=0$. For $e^2<0$, however, there is an initial minimum
at $N=0$ beyond which $E$ rises $\sim{N}$ until a maximum is reached at $N={N}_{crit}$,
\begin{eqnarray}
{N}_{crit}\sim\frac{T}{V{|e|}^{2}}\sim\frac{1}{e^2}.
\end{eqnarray}
Beyond this point $E$ decreases as $-N^2$. There is no {\it stable} minimum.  
One infers that the divergent nature of the perturbation series emerges when more than $N_c$
terms are considered. For $n<N_{crit}$ the ${f}_{n}{e}^{2n}$ terms decrease and ${f}_{n}{e}^{2n}\sim{f}_{n+1}{e}^{2n+2}$ at $n\sim{N}_{crit}$.
\begin{eqnarray}
&&\frac{f_{n+1}}{f_{n}}\sim\frac{1}{e^{2}}\sim N_{crit}\sim
n\nonumber,\\[10pt]
&&\qquad\Rightarrow f_{n}\sim n!.
\end{eqnarray}
\subsection{Asymptotic Series/Borel summation}
Consider a function $f(g)$ expressed in terms of a power series expansion
\begin{eqnarray}
f(g)=\sum_{n=0}^{\infty}f_{n}g^{n}.
\end{eqnarray}
Consider a domain $\cal{D}$ of the complex $g$-plane such that $arg\;g<\pi/2$. The series
is said to be asymptotic inside $\cal{D}$ if the series diverges for all $g\ne{0}$ and
\begin{eqnarray}
\qquad\qquad\quad\;\bigg{|}f(g)-\sum_{n=0}^{N}f_{n}g^{n}\bigg{|}\leq\;f_{N+1}|g|^{N+1}.
\end{eqnarray}
The crucial property of asymptotic series is that the error made in truncating
the series is less than the first neglected term. If $f_n\sim n!$ then successive terms will decrease
until a minimum is reached at $N_{opt}$ terms, thereafter the size of the terms will increase without
limit. By truncating the series at $n=N_{opt}$ the best possible approximation is found.\\

One can define a function which is asymptotic to the series by using the
Borel method. If $f_{n}\sim{n}!$ one defines the Borel transform of the series
\begin{eqnarray}
B[f](z)=\sum_{n=0}^{\infty}\frac{f_n}{n!}g^n.
\end{eqnarray}
This series will now have a finite radius of convergence. One can then
write
\begin{eqnarray}
f(g)=\int_{0}^{\infty}{dz}{e}^{-z/g}\sum_{n=0}^{\infty}\frac{f_n}{n!}z^n.
\end{eqnarray}
This follows since if one performs the integral term by term and uses the result 
\begin{eqnarray}
\int_{0}^{\infty}{dz}{e}^{-z/g}z^n={n}!g^n,
\end{eqnarray}
one formally reproduces the divergent power series for $f(g)$. If this
series has a finite radius of convergence then one can show that the $B[f](z)$ series has
infinite radius of convergence , and $f(g)$ is equal to the Borel sum
\begin{eqnarray}
f(g)=\int_{o}^{\infty}{dz}{e}^{-z/g}B[f](z).
\end{eqnarray}
If $f(g)$ has zero radius of convergence with $f_n\sim{n}!$ then $B[f](z)$ will have
finite radius of convergence, and can be {\it analytically continued} outside of that radius to
the whole of the integration range $[0,\infty]$
\begin{eqnarray}
f(g)\approx \int_{0}^{\infty}{dz}{e}^{-z/g}B[f](z).
\end{eqnarray}
Here the $\approx$ symbol means ``is asymptotic to''. Notice that 
there is in general not a {\it unique} function to which the series is asymptotic
since we can always add a term ${e}^{-C/g}$, which has an identically zero Taylor expansion
in powers of $g$, and the series will also be asymptotic to that function. If we have
information on the analytic structure of $f(g)$ in the complex $g$-plane it is sometimes
possible to enforce that $C=0$ (Watson's Theorem).\\

Two relevant examples which will be important later when we introduce renormalons
are alternating and fixed-sign factorial growth.
First consider the alternating factorial series,
\begin{eqnarray}
{f}_{-}(g)=\sum_{n=0}^{\infty}{(-1)}^{n}n!g^n. 
\end{eqnarray}
This has the Borel transform
\begin{eqnarray}
B[f_{-}](z)=\sum_{n=0}^{\infty}{(-1)}^{n}{z}^{n}.
\end{eqnarray}
This series converges for $|z|<1$, and may be analytically continued to $B[{f}_{-}](z)=1/(1+z)$ on
the whole range $[0,\infty]$ one then finds,
\begin{eqnarray}
{f}_{-}(z)\approx\int_{0}^{\infty}{dz}{e}^{-z/g}\frac{1}{(1+z)}=-{\rm{Ei}}(-1/g).
\end{eqnarray}
Here ${\rm{Ei}}(x)$ is the Exponential integral function defined (for $x<0$) as
\begin{eqnarray}
{\rm{Ei}}(x)=-\int_{-x}^{\infty}{dt}\frac{e^{-t}}{t}
\end{eqnarray}

Now consider the factorial series
\begin{eqnarray}
{f}_{+}(g)=\sum_{n=0}^{\infty}n!{g}^{n}
\end{eqnarray}
This has the Borel transform
\begin{eqnarray}
B[f_+](z)=\sum_{n=0}^{\infty}{z}^{n}.
\end{eqnarray}
This series converges for $|z|<1$, and may be analytically continued to $B[f_+](z)=1/(1-z)$,
\begin{eqnarray}
{f}_{+}(g)=\int_{0}^{\infty}{dz}{e}^{-z/g}\frac{1}{(1-z)}={\rm{Ei}}(1/g).
\end{eqnarray}
For $x>0$ ${\rm{Ei}}(x)$ is defined by
\begin{eqnarray}
{\rm{Ei}}(x)=-PV\;\int_{-x}^{\infty}{dt}\frac{e^{-t}}{t}.
\end{eqnarray}
There is a pole at $z=1$ in the Borel plane which renders the result ambiguous,
with a $\pm{i\pi}$ contribution depending on whether the integration contour passes
above or below the pole. Use of the principal value (PV) prescription is
equivalent to averaging over these choices.

\subsection{Proliferation of Feynman diagrams and Instantons}
Consider the following integral expressed in terms of a power series expansion
\begin{eqnarray}
I(g)&=&\frac{1}{\sqrt{2\pi}}\int_{-\infty}^{+\infty}dx\;e^{-(x^{2}/2+gx^{4}/4)}
\nonumber \\[10pt]
&=&\sum_{n=0}^{\infty}I_{n}g^{n}.
\end{eqnarray}
This integral is the generating function for the number of Feynman diagrams
contributing to the vacuum-to-vacuum transition amplitude of $\phi^4$ field theory. One has
\begin{eqnarray}
I_{n}&=&\frac{1}{\sqrt{2\pi}}\frac{(-1)^{n}}{4^{n}n!}\int_{-\infty}^{+\infty}x^{4n}e^{-x^{2}/2}dx\nonumber\\[10pt]
&=&\frac{(-1)^{n}}{\sqrt{\pi}}\frac{\Gamma(2n+1/2)}{\Gamma(n+1)},\nonumber \\[10pt]
I_{n}&\sim&\frac{(-4)^{n}}{2\pi}(n-1)!  
\end{eqnarray}
and so the combinatoric growth in the number of Feynman Diagrams would be expected,
other things being equal, to contribute to factorial growth of the coefficients.\\

We now turn to a brief discussion of instantons \cite{r10}.
Consider a generic Green function ${\cal{G}}(g)$ for a simplistic field
theory of a single field at a single spacetime point. 
\begin{eqnarray}
\mathcal{G}(g)=\frac{1}{g}\int_{-\infty}^{+\infty}d\phi\;e^{-S(\phi)/g}.
\end{eqnarray}
This expression can be written in terms of the Borel transform of
$\mathcal{G}(g)$.
\begin{eqnarray}
{\mathcal{G}}(g)&=&\frac{1}{g}\int_{0}^{\infty}e^{-z/g}\;B[\mathcal{G}](z)dz.
\end{eqnarray}
By inspection, the Borel transform of $\mathcal{G}(g)$ is found to be
\begin{eqnarray}
B[\mathcal{G}](z)=\int_{-\infty}^{+\infty}d\phi\;\delta(z-S(\phi)),
\end{eqnarray}
and this can be rewritten by change of integration variable as,
\begin{eqnarray}
B[\mathcal{G}](z)&=&\int_{-\infty}^{+\infty}dS(\phi)\Bigg{[}\frac{\partial
S(\phi)}{\partial\phi}\Bigg{]}^{-1}\delta(z-S(\phi))\nonumber\\
&=&\sum_{i}\Bigg{[}\frac{\partial
S(\phi)}{\partial\phi}\Bigg{]}^{-1}\Bigg{|}_{\phi=\phi_{i}}.
\end{eqnarray}
Here $\phi_{i}$ label all solutions $\phi$ obeying $S(\phi)=z$.
We can see that singularities in the Borel plane occur at values of
$S[\phi]$ for which $\phi$ satisfies
\begin{eqnarray}
\frac{\partial S(\phi)}{\partial\phi}=0. 
\end{eqnarray}
Hence $\phi_{i}$ represent extrema of the action and they are
therefore solutions to the classical equations of motion.
Consequently, singularities exist in the Borel plane at values of
the action corresponding to these classical solutions. We expand
the action in the region of $\phi_i$,
\begin{eqnarray}
S(\phi)&\simeq&S(\phi_{i})+\frac{1}{2}S''(\phi_{i})(\phi-\phi_{i})^{2}.
\end{eqnarray}
We then rearrange this whilst defining $z_{i}=S(\phi_{i})$.
\begin{eqnarray}
\phi-\phi_{i}&\simeq& \Bigg{(}2\frac{(S(\phi)-S(\phi_{i}))}{S''(\phi_{i})}\Bigg{)}^{1/2},\nonumber \\
\Rightarrow\Bigg{[}\frac{\partial
S(\phi)}{\partial\phi}\Bigg{]}^{-1}&\simeq&\frac{1}{\sqrt{S''(\phi_{i})}}\frac{1}{\sqrt{z-z_{i}}}.
\end{eqnarray}
Substituting this result back into the Borel integral of Eq.(56) one finds
\begin{eqnarray}
\mathcal{G}(g)&=&\sum_{n=0}^{\infty}\mathcal{G}_{n}g^{n},\nonumber \\
&\simeq&\int_{0}^{\infty}e^{-z/g}(1-z/z_{i})^{-1/2}dz.
\end{eqnarray}
The series coefficients are then determined to be
\begin{eqnarray}
\mathcal{G}_{n}&\simeq&\Bigg{(}\frac{1}{z_{i}}\Bigg{)}^{n}\frac{\Gamma(n+1/2)}{\Gamma(1/2)}\nonumber,\\
&\sim&\Bigg{(}\frac{1}{z_{i}}\Bigg{)}^{n}\frac{n!}{n^{1/2}}.
\end{eqnarray}
Thus one finds $n!$ growth in perturbative coefficients. The
Borel plane singularities are at positions in the $z$-plane corresponding to the
values of the action representing instanton solutions. By considering specific examples
such as the anharmonic oscillator \cite{r10} one can make a concrete connection between this $n!$ instanton
contribution and that due to the proliferation of Feynman diagrams discussed above.

\section{Large-$N_f$ approximation for vacuum polarization}
Consider the Adler $D$-function we discussed earlier (Eq.(30)) with perturbative
expansion. The $n$th perturbative coefficient coefficient $d_n$ in Eq.(31) may be expanded in powers of $N_f$ the number of quark flavours
\begin{eqnarray}
{d}_{n}={d}_{n}^{[n]}{N}_{f}^{{n}}+{d}_{n}^{[n-1]}{N}_{f}^{n-1}+\ldots+{d}_{n}^{[0]}.
\end{eqnarray}
The leading large-$N_f$ coefficient ${d}_{n}^{[n]}$ may be evaluated to all-orders
since it derives from a 
restricted set of diagrams obtained by inserting a chain of fermion bubbles 
inside the quark loop (see Fig.2).
\begin{figure*}[t]
\centering
\begin{tabular}{lr}
\psfrag{q}{$q$}\psfrag{k}{\small{\mbox{$k$}}}\includegraphics[width=0.45\textwidth]{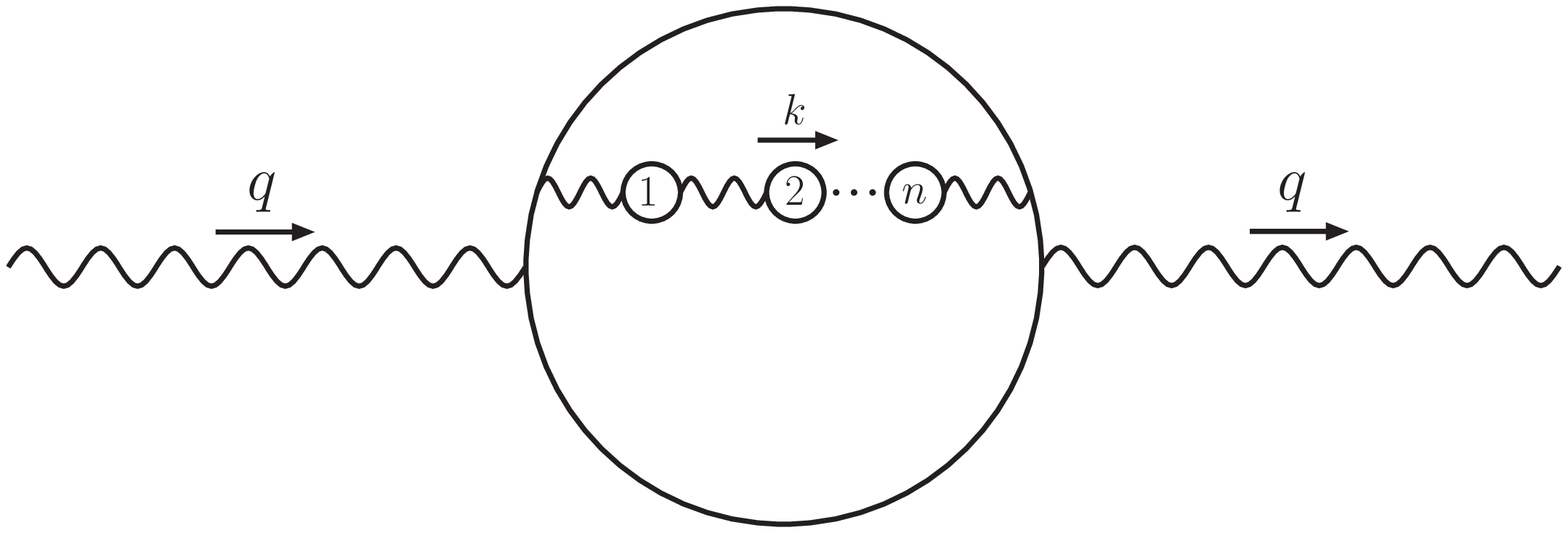}\hspace{.05\textwidth}&\hspace{.05\textwidth}\psfrag{q}{$q$}\psfrag{k}{\small{\mbox{$k$}}}\includegraphics[width=0.45\textwidth]{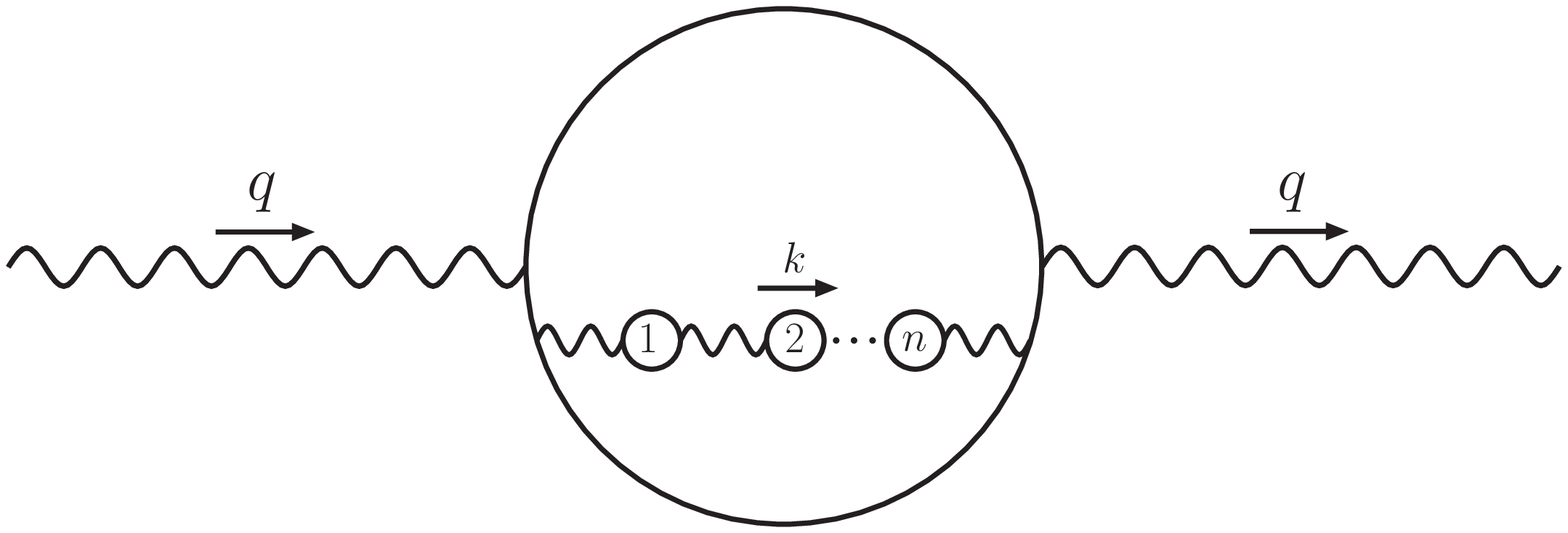}\\\\
\multicolumn{2}{c}{\psfrag{q}{$q$}\psfrag{k}{$k$}\includegraphics[width=0.45\textwidth]{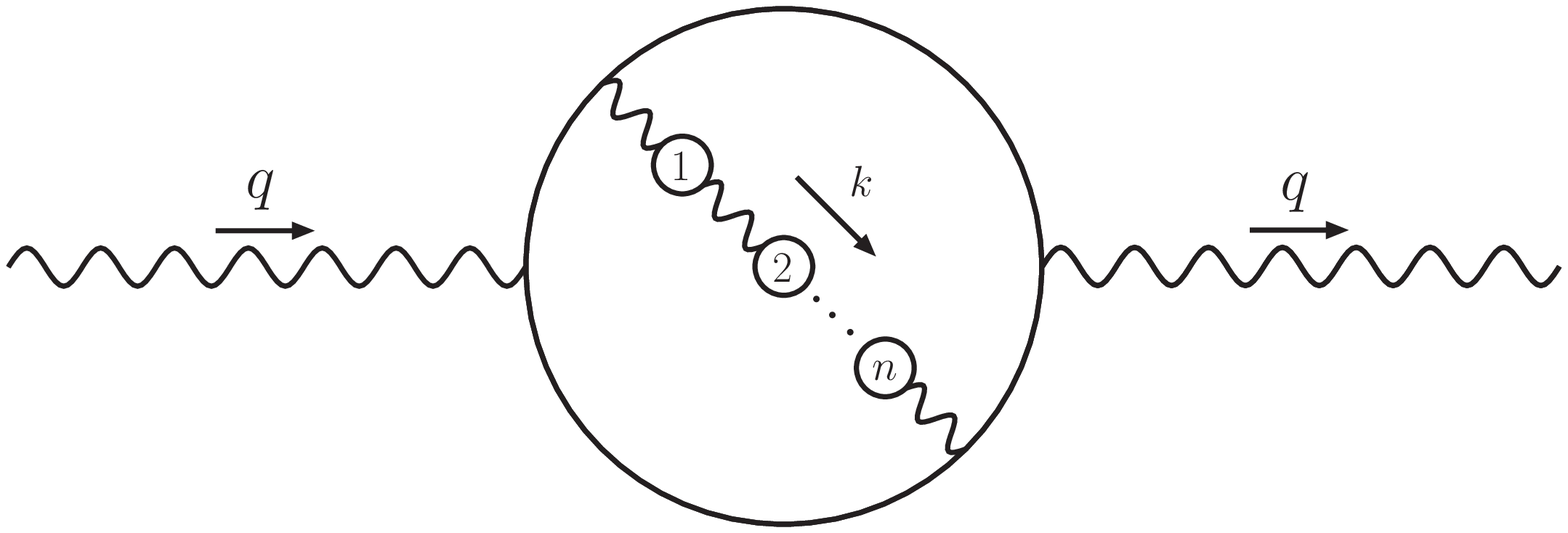}}\\
\end{tabular}
\caption{Bubble-chain diagrams contributing to $d_n^{[n]}$.}
\end{figure*}
A crucial ingredient is the chain of $n$-bubbles, ${B}_{(n)}^{\mu\nu}(k^2)$.
This may be defined as a product of bubbles ${\Pi}_{\mu\nu}(k^2)$ and propagators ${P}_{\mu\nu}(k^2)$ 
\begin{eqnarray}
-i{\Pi}_{\mu\nu}(k^2)&=&-i(k^2{g}_{\mu\nu}-{k}_{\mu}{k}_{\nu}){\Pi}_{0}(k^2)
\nonumber \\
-i{P}_{\mu\nu}&=&-i\frac{({g}_{\mu\nu}-k^2(1-\xi){k}_{\mu}{k}_{\nu})}{k^2}.
\end{eqnarray}
Stringing the bubbles and propagators together gives
\begin{eqnarray}
{B}_{(n)}^{\mu\nu}(k^2)=\prod_{k=1}^{n}\Bigg{[}(-iP^{\alpha_{k}\beta_{k}})(-i\Pi_{\beta_{k}\alpha_{k+1}})\Bigg{]}(-iP^{\alpha_{n+1}\nu}).
\end{eqnarray}
Here $\mu={\alpha}_{1}$. To evaluate this we will need the results

\begin{eqnarray}
P^{\alpha_{k}\beta_{k}}\Pi_{\beta_{k}\alpha_{k+1}}=\frac{1}{k^{2}}\Pi^{\alpha_{k}}_{\;\;\;\;\alpha_{k+1}}\qquad
\end{eqnarray}
and
\begin{eqnarray}
\prod_{k=1}^{n}\Bigg{[}\Pi^{\alpha_{k}}_{\;\;\;\;\alpha_{k+1}}\Bigg{]}=
\Pi^{\alpha_{1}}_{\;\;\;\;\alpha_{n+1}}\Pi^{n-1}_{0}(k^{2})^{n-1}.
\end{eqnarray}
We finally obtain
\begin{eqnarray}
B_{(n)}^{\mu\nu}(k^{2})&=&\frac{(-1)^{n}}{k^{2}}\Pi^{n-1}_{0}
\Pi^{\alpha_{1}}_{\;\;\;\;\alpha_{n+1}}(-iP^{\alpha_{n+1}\nu})\nonumber\\[10pt]
&=&(-1)^{n}(\Pi_{0})^{n}\Bigg{(}\frac{-i}{k^{2}}\Bigg{)}
\Bigg{[}g^{\mu\nu}-\frac{k^{\mu}k^{\nu}}{k^{2}}\Bigg{]},\nonumber \\[10pt]
B_{(n)}^{\mu\nu}(k^{2})&=&(-1)^{n}(\Pi_{0})^{n}[-iP^{\mu\nu}(k^{2},\xi=0)].
\end{eqnarray}
This is proportional to the Landau gauge ($\xi=0$) propagator. Explicitly
\begin{eqnarray}
{B}_{(n)}^{\mu\nu}=\frac{(k^2{g}^{\mu\nu}-{k}_{\mu}{k}_{\nu})}{{(k^2)}^{2}}{\left[-\frac{N_f}{3}\left(\ln\frac{k^2}{\mu^2}+C\right)\right]}^{n}.
\end{eqnarray}
The constant $C$ depends on the subtraction procedure used to renormalise the bubble. With
$\overline{MS}$ subtraction $C=-\frac{5}{3}$. We shall choose to work in the ``V-scheme'' which corresponds
to $\overline{MS}$ with the scale choice $\mu^2={e}^{-5/3}Q^2$, in which case $C=0$.
Applying the Feynman rules to the three diagrams then gives ${d}_{n}^{[n]}{a}^{n+1}$
\begin{eqnarray}
&\sim&a\int\frac{d^{4}k}{(2\pi)^{4}}\frac{d^{4}p}{(2\pi)^{4}}
\nonumber \\
&\Bigg{[}&B_{(n)}^{\sigma\rho}(k^{2})\textrm{Tr}\Bigg{(}\gamma_{\nu}\frac{1}{p\slsh+q\slsh+k\slsh}
\gamma_{\rho}\frac{1}{p\slsh+q\slsh}\gamma_{\mu}\frac{1}{p\slsh}
\gamma_{\sigma}\frac{1}{p\slsh+k\slsh}\Bigg{)}\nonumber\\[10pt]
&+&2B_{(n)}^{\sigma\rho}(k^{2})\textrm{Tr}\Bigg{(}\gamma_{\nu}\frac{1}{p\slsh+q\slsh}
\gamma_{\mu}\frac{1}{p\slsh}\gamma_{\sigma}\frac{1}{p\slsh+k\slsh}
\gamma_{\rho}\frac{1}{p\slsh}\Bigg{)} \Bigg{]}.
\end{eqnarray}
The loop integrals can be evaluated using the Gegenbauer polynomial $x$-space technique, with the result \cite{r11,r12}
\begin{eqnarray}
&{d}_{n}^{[n]}(V)&=\frac{-2}{3}(n+1){\left(\frac{-1}{6}\right)}^{n}\left[-2n-\frac{n+6}{{2}^{n+2}}\right.
  \nonumber \\
&+&\frac{16}{n+1}{\sum_{\frac{n}{2}+1>m>0}}m(1-{2}^{-2m})
\nonumber \\
&\times&\left. (1-{2}^{2m-n-2}){\zeta}_{2m+1}\right]{n}!\;.
\end{eqnarray}
\subsection{Leading-$b$ approximation and QCD renormalons}
The large-$N_f$ result of the last section can describe QED vacuum polarization, but for QCD the corrections
to the gluon propagator involve gluon and ghost loops, and are gauge $(\xi)$-dependent. The result
for ${\Pi}_{0}(k^2)$ is proportional to $-N_f/3$ which is the first QED beta-function coefficient, $b$. 
In QCD one expects large-order behaviour of the form ${d}_{n}\sim K{n}^{\gamma}{(b/2)}^{n}n!$ \cite{r2} involving
the QCD beta-function coefficient $b=(33-2N_f)/6$, it is then natural to relace $N_f$ by $(33/2-3b)$ to
obtain an expansion in powers of $b$ \cite{r13,r14,r15}
\begin{eqnarray}
{d}_{n}={d}_{n}^{(n)}{b}^{n}+{d}_{n}^{(n-1)}{b}^{n-1}+\ldots+{d}_{n}^{(0)}.
\end{eqnarray}
The leading-$b$ term ${d}_{n}^{(L)}\equiv{d}_{n}^{(n)}b^n={(-3)}^{n}{d}_{n}^{[n]}b^n$ can
then be used to approximate $d_n$ to all-orders, and an all-orders resummation of these results
performed to obtain ${\cal{D}}^{(L)}_{PT}(Q^2)$. In order for the result to be renormalisation
scheme independent we need to use the one-loop form of the coupling $a(Q^2)=2/b{\ln}(Q^2/{\Lambda}^2$,
and in what follows we will work in the $V$-scheme discussed earlier so that $C=0$ in Eq.(70). 
If we use the Borel method to define the all-orders perturbative result we obtain
\begin{eqnarray}
{\cal{D}}^{(L)}_{PT}({Q}^{2})={\int_{0}^{\infty}}{dz}\,{e}^{-z/a({Q}^{2})}B[{\cal{D}}^{(L)}_{PT}](z)\;.
\end{eqnarray}
The Borel transform is given by \cite{r14}
\begin{eqnarray}
B[{\MD}^{(L)}_{PT}](z)
&=&\sum_{n=1}^{\infty}\frac{A_{0}(n)-A_{1}(n)z_{n}}
{\Big{(}1+\frac{z}{z_{n}}\Big{)}^{2}}+\frac{A_{1}(n)z_{n}}
{\Big{(}1+\frac{z}{z_{n}}\Big{)}}\nonumber\\
&+&\sum_{n=1}^{\infty}\frac{B_{0}(n)+B_{1}(n)z_{n}}
{\Big{(}1-\frac{z}{z_{n}}\Big{)}^{2}}-\frac{B_{1}(n)z_{n}}
{\Big{(}1-\frac{z}{z_{n}}\Big{)}}\label{BorelD}.
\end{eqnarray}
The residues are given by
\begin{eqnarray}
A_{0}(n)&=&\frac{8}{3}\frac{(-1)^{n+1}(3n^{2}+6n+2)}{n^{2}(n+1)^{2}(n+2)^{2}}
\nonumber \\
A_{1}(n)&=&\frac{8}{3}\frac{b(-1)^{n+1}(n+\frac{3}{2})}{n^{2}(n+1)^{2}(n+2)^{2}}\nonumber\\[10pt]
\nonumber
\end{eqnarray}
\begin{eqnarray}
&&B_{0}(1)=0,\quad B_{0}(2)=1,\quad B_{0}(n)=-A_{0}(-n)\quad
n\geq3\nonumber \\
&&B_{1}(1)=0,\quad B_{1}(2)=-\frac{b}{4},\quad B_{1}(n)=-A_{1}(-n)\quad
n\geq3.
\end{eqnarray}
\begin{figure*}[t]
\centering
{\epsfig{file = 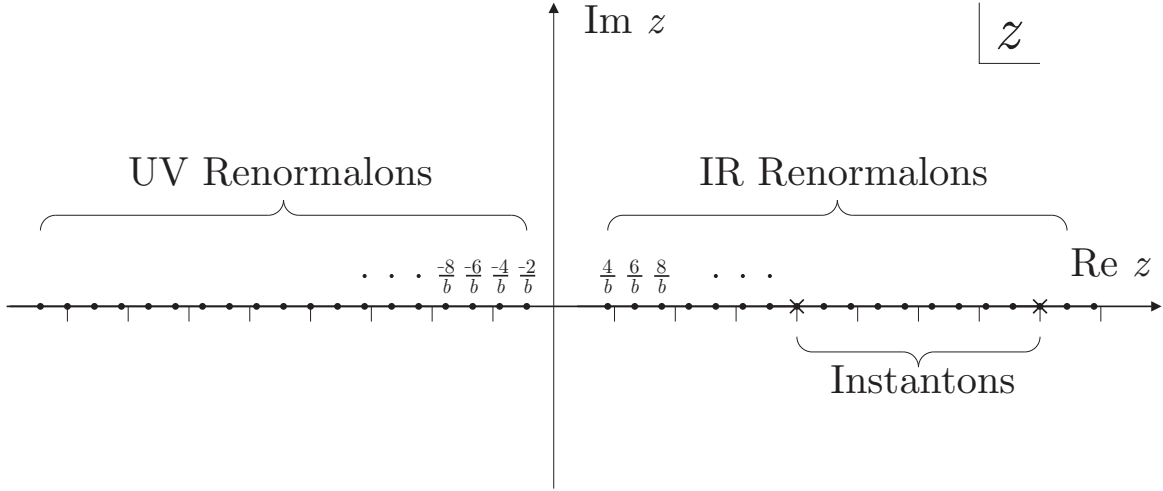, width = 15.38cm}}
\caption{Borel plane singularities for the Adler $D$-function. $N_f=5$ active quark flavours are assumed.}
\end{figure*}
In Fig.3 we show the Borel plane singularities. For the Adler function in leading-$b$ approximation there are single and double
poles in $B[{\MD}^{(L)}_{PT}](z)$ at positions $z=z_n$ and $z=-z_n$ with $z_n\equiv{2n}/b$, $n=1,2,3$.
The singularities on the positive real semi-axis are the {\it infrared} renormalons, ${IR}_{n}$ and those on the
negative real semi-axis are {\it ultraviolet} renormalons, ${UV}_{n}$. We shall see that they correspond to
integration over the bubble-chain momentum $k^2$ in the regions $k^2<Q^2$ and $k^2>Q^2$, respectively.
The ${IR}_{1}$ singularity is absent for $\MD$ for reasons which we shall shortly discuss. Hence the singularity
nearest the origin is the ${UV}_{1}$ renormalon, which generates the leading asymptotic behaviour
\begin{eqnarray}
{d}_{n}^{(L)}\approx\frac{(12n+22)}{27}n!{\left(\frac{-b}{2}\right)}^{n}.
\nonumber
\end{eqnarray}
The ${IR}_{n}$ singularities lie on the integration contour, and hence there is
an amiguous imaginary part depending on whether the contour is routed above or below the pole.
This is structurally the same as terms in the non-perturbative Operator Product Expansion (OPE).
We see that there are also singularities corresponding to instanton contributions. 
These are at positions $z_n=4n$ corresponding to the $nI{\bar{I}}$ pair contribution. The leading
$I{\bar{I}}$ instanton singularity at $z=4$ lies well to the right of the leading $IR_2$ singularity and
so does not dominate the asymptotic behaviour.
\subsection{OPE and IR renormalon ambiguities}
The regular OPE is a sum over the contributions of condensates with different mass dimensions.
In the case of the Adler function the dimension four gluon condensate is the leading contribution
\begin{eqnarray}
{G}_{0}(a(Q^2))=\frac{1}{Q^4}\langle 0|GG|0\rangle{C}_{GG}(a(Q^2))\;,  
\end{eqnarray}
where ${C}_{GG}(a(Q^2))$ is the Wilson coefficient. The OPE is of the form
\begin{eqnarray}
{\MD}_{NP}(Q^2)=\sum_{n}{\cal{C}}_{n}{\left(\frac{\Lambda^2}{Q^2}\right)}^{n}
\end{eqnarray}
The ${n}^{\rm{th}}$ term in this expansion will have the structure
\begin{eqnarray}
{\cal{C}}_{n}(a(Q^2))={C}_{n}{[a(Q^2)]}^{{\delta}_{n}}(1+O(a))\;.
\end{eqnarray}
The exponent $\delta_n$ corresponding to the anomalous dimension of the condensate operator
concerned. 
Non-logarithmic UV divergences lead to an ambiguous imaginary part in the
coefficient \cite{r16} so that ${C}_{n}={C}_{n}^{(R)}\pm i{C}_{n}^{(I)}$. If one considers an ${IR}_{n}$
renormalon singularity in the Borel plane to be of the form $K_n/{(1-z/{z}_{n})}^{{\gamma}_{n}}$
then one finds an ambiguous imaginary part arising of the form
\begin{eqnarray}
Im [{\cal{D}}_{PT}]=\pm{K}_{n}\frac{\pi{z}_{n}^{{\gamma}_{n}}}{\Gamma({\gamma}_{n})}{e}^{-{z}_{n}/a(Q^2)}{a}^{1-{\gamma}_{n}}[1+O(a)]\;.
\end{eqnarray}
Here the $\pm$ ambiguity comes from routing the contour above or below the real $z$-axis in the
Borel plane. This is structurally the same as the ambiguous OPE term above, and if
${C}_{n}^{(I)}=K_{n}\pi{z}_{n}^{{\gamma}_{n}}/\Gamma({\gamma}_{n})$ and ${\delta}_{n}=1-{\gamma}_{n}$, then the PT Borel and
NP OPE ambiguities can potentially cancel against each other \cite{r17}. Taking a PV of the Borel integral corresponds
to averaging over the $\pm$ possibilities. Notice that there is no condensate of dimension two in the OPE,
the leading gluon condensate being of dimension four. This explains the absence of the $IR_1$ singularity
at $z=2/b$ evident in Fig.3.

\begin{figure*}[t]
\centering
\begin{tabular}{lcr}
\includegraphics[width=0.3\textwidth]{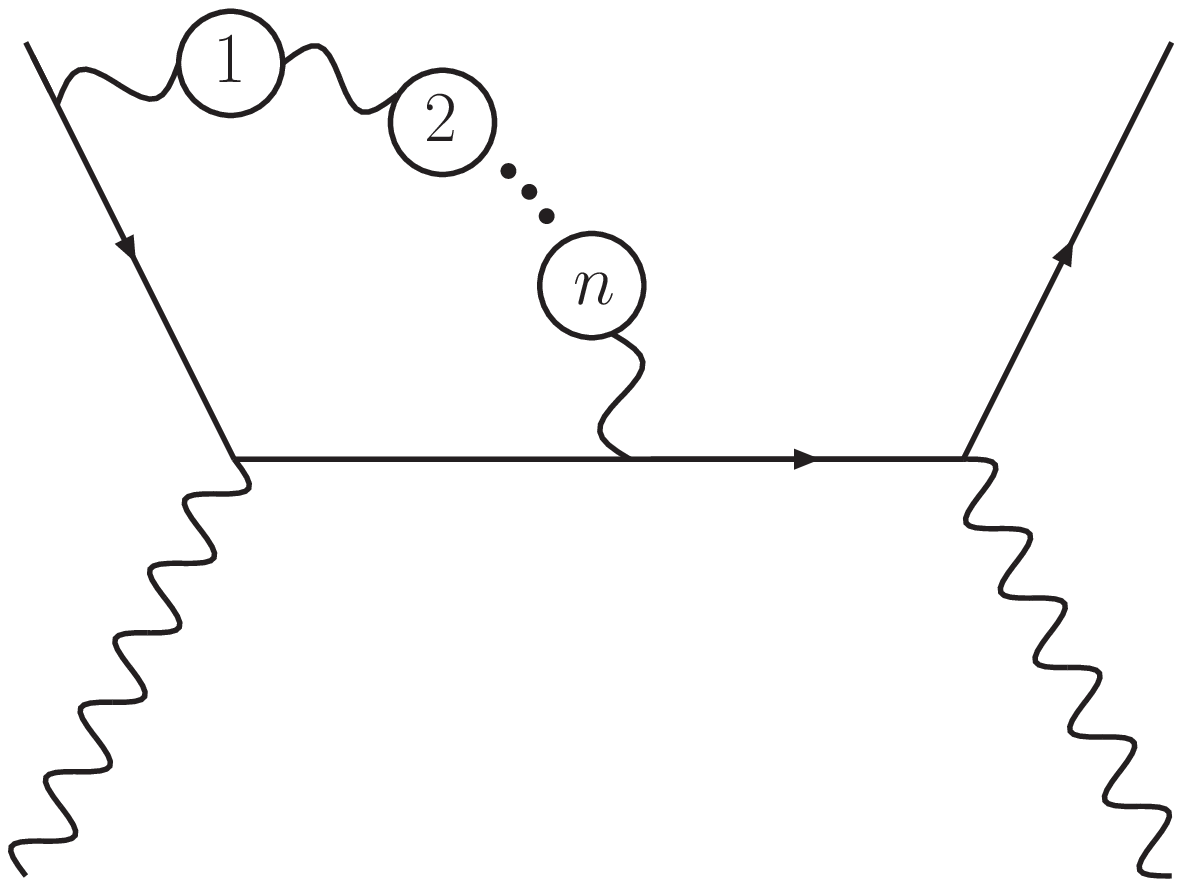}&\includegraphics[width=0.3\textwidth]{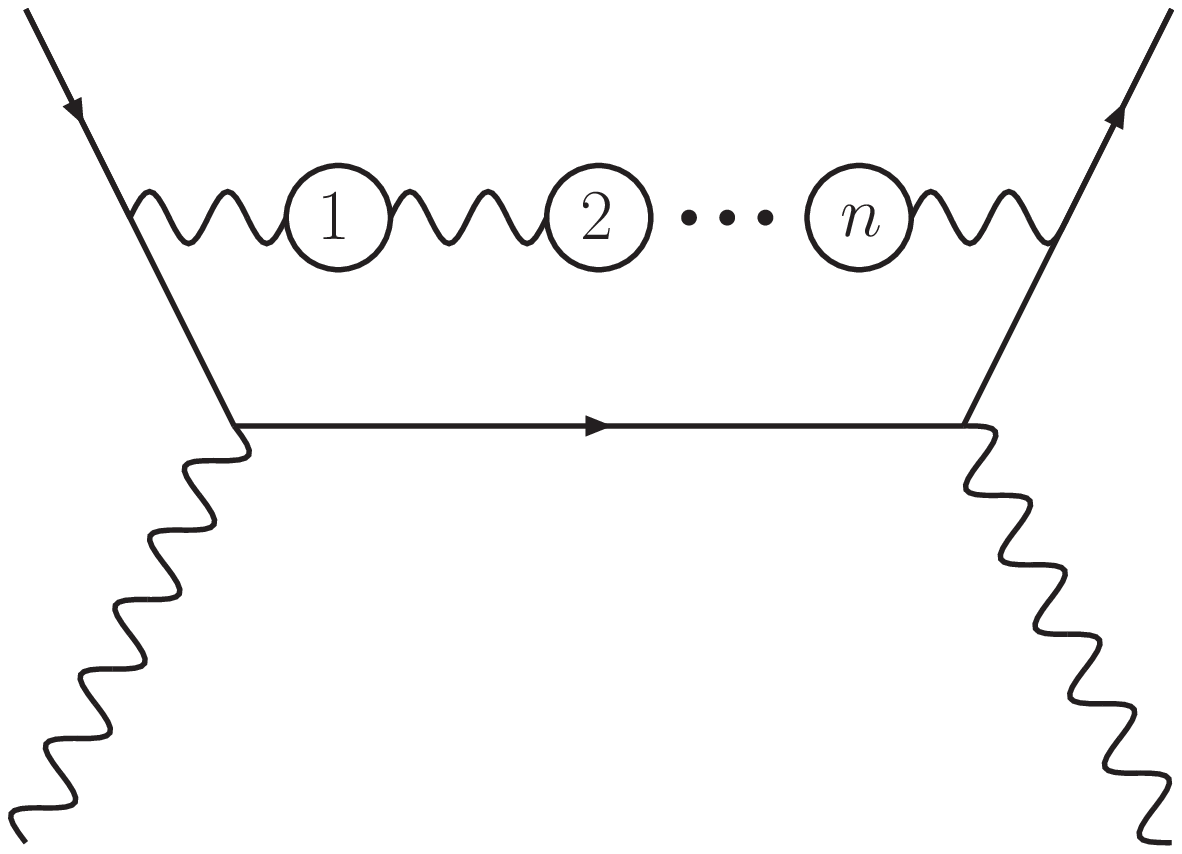}&
{\includegraphics[width=0.3\textwidth]{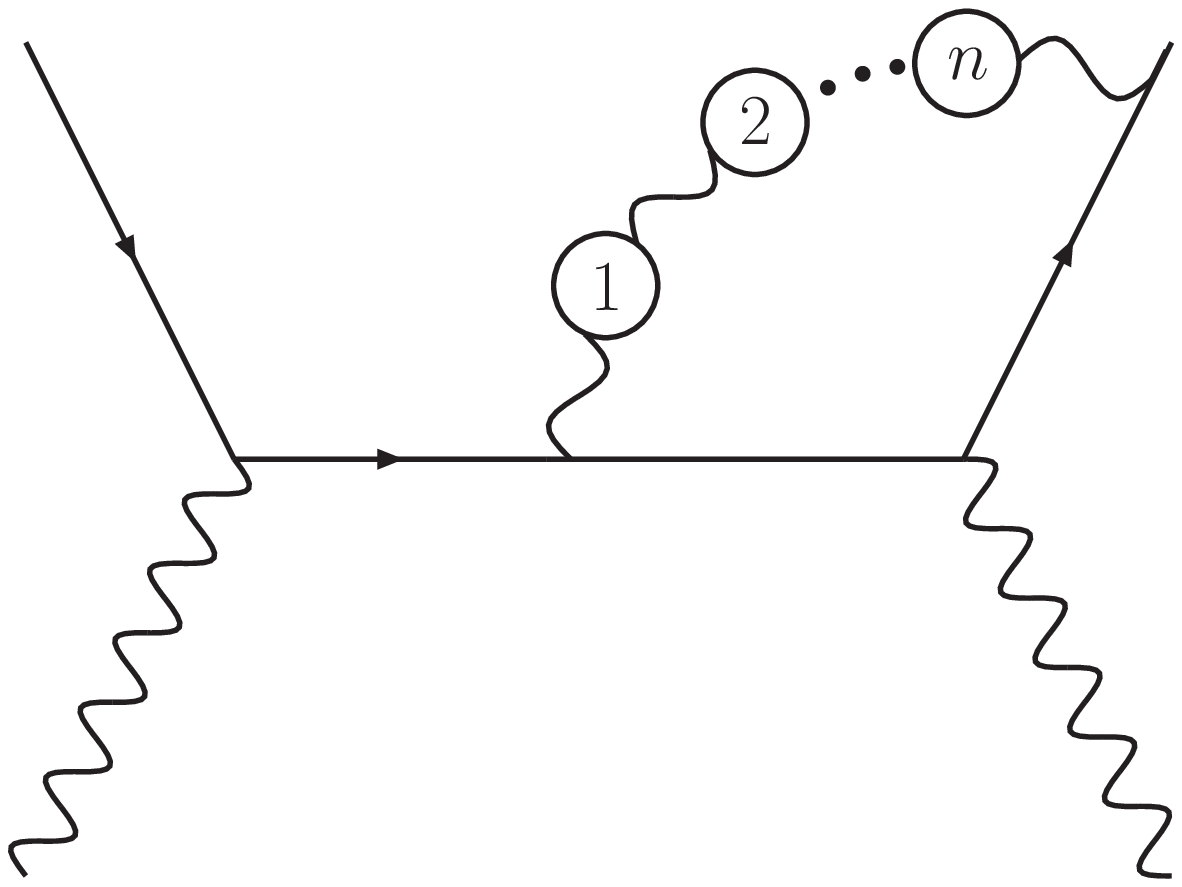}}\\
\end{tabular}
\caption{Large-$N_f$ contributions to the DIS sum rules at $n$th order in perturbation theory}
\end{figure*}

\section{Freezing of Euclidean observables/skeleton expansion}
We wish to investigate the $Q^2$-dependence of Euclidean observables such as the Adler function, and
in particular whether ${\MD}^{(L)}(Q^2)={\MD}^{(L)}_{PT}(Q^2)+{\MD}^{(L)}_{NP}(Q^2)$ can remain
finite as $Q^2\rightarrow{0}$. We shall find that these components in the one-chain QCD skeleton expansion
both vanish as $Q^2\rightarrow{0}$ \cite{r1}. Let us introduce two other observables.
The polarised Bjorken (pBj) and GLS sum rules are
defined as \cite{r18,r19}
\begin{eqnarray}
K_{pBj}&\equiv&\int_{0}^{1}g_{1}^{ep-en}(x,Q^{2})dx\nonumber \\
&=&\frac{1}{6}\Bigg{|}\frac{g_{A}}{g_{V}}\Bigg{|}\Bigg{(}1-\frac{3}{4}C_{F}\MK(Q^2)\Bigg{)}\;,
\end{eqnarray}
\begin{eqnarray}
K_{GLS}&\equiv&\frac{1}{2}\int_{0}^{1}F_{3}^{\bar{\nu}p+\nu p}
(x,Q^{2})dx\nonumber \\
&=&3\Bigg{(}1-\frac{3}{4}C_{F}\MK(Q^2)\Bigg{)}\label{Kg}\;.
\end{eqnarray}
$\MK(Q^2)$ being the QCD corrections  to the parton model
result.
 We have neglected contributions due to
``light-by-light'' diagrams which when omitted render the
perturbative corrections to $K_{GLS}$ and $K_{pBj}$ identical.
Finally, the unpolarised Bjorken sum rule (uBj) is defined as \cite{r20}
\begin{eqnarray}
U_{uBj}&\equiv&\int_{0}^{1}F_{1}^{\bar{\nu}p-\nu p}(x,Q^{2})dx \nonumber \\
&=&\Bigg{(}1-\frac{1}{2}C_{F}\MU(Q^2)\Bigg{)}.
\end{eqnarray}
Leading-$b$ results for ${\MK}^{(L)}_{PT}(Q^2)$ and ${\MU}^{(L)}_{PT}(Q^2)$ can be computed from the diagrams in Fig.4.
The expressions for $B[{\MK}^{(L)}_{PT}](z)$ and $B[{\MU}^{(L)}_{PT}](z)$ are \cite{r21,r22}
\begin{eqnarray}
B[{\cal{K}}^{(L)}_{PT}](z)&=&\frac{4/9}{\Big{(}1+\frac{z}{z_{1}}\Big{)}}-\frac{1/18}{\Big{(}1+\frac{z}{z_{2}}\Big{)}}+\frac{8/9}{\Big{(}1-\frac{z}{z_{1}}\Big{)}}
\nonumber \\
&-&\frac{5/18}{\Big{(}1-
\frac{z}{z_{2}}\Big{)}} \\
B[{\cal{U}}^{(L)}_{PT}](z)&=&\frac{1/6}{\Big{(}1+\frac{z}{z_{2}}\Big{)}}+\frac{4/3}{\Big{(}1-\frac{z}{z_{1}}\Big{)}}
\nonumber \\
&-&\frac{1/2}{\Big{(}1-\frac{z}{z_{2}}\Big{)}}\;.
\end{eqnarray}
These expressions are significantly simpler than for the Adler function since one is inserting
the bubble chain in a tree-level diagram, rather than a one-loop one. There are a finite number of simple poles.
Using the integrals
\begin{eqnarray}
\int_{0}^{\infty}\frac{e^{-z/a}}{(1+z/z_{n})}=-z_{n}e^{z_{n}/a}\textrm{Ei}(-z_{n}/a)\label{BI1} \nonumber \\
\int_{0}^{\infty}\frac{e^{-z/a}}{(1+z/z_{n})^{2}}=z_{n}\Bigg{[}1+\frac{z_{n}}{a}e^{-z_{n}/a}\textrm{Ei}(z_{n}/a)\Bigg{]},
\end{eqnarray}
we can obtain the following resummed expressions \cite{r14}
\begin{eqnarray}
{\MD}^{(L)}_{PT}(Q^2)&=&\sum_{n=1}^{\infty}[{z}_{n}{e}^{{z}_{n}/a(Q^2)}{\rm{Ei}}\left(\frac{z_n}{a(Q^2)}\right)
\nonumber \\
&\times&\left[\frac{z_n}{a(Q^2)}(A_0(n)-z_n{A}_{1}(n))-z_n{A}_{1}(n)\right]
\nonumber \\
&+&(A_0(n)-{z}_{n}A_1(n))]
\nonumber \\
&+&\sum_{n=1}^{\infty}{z}_{n}[{e}^{-z_n/a(Q^2)}{\rm{Ei}}\left(\frac{z_n}{a(Q^2)}\right)
\nonumber \\
&\times&\left[\frac{z_n}{a(Q^2)}(B_0(n)+{z}_{n}{B}_{1}(n))-{z}_{n}{B}_{1}(n)\right]
\nonumber \\
&-&({B}_{0}(n)+z_n{B}_{1}(n))],
\end{eqnarray}
\begin{eqnarray}
{\MK}^{(L)}_{PT}(Q^2)&=&\frac{1}{9b} \Bigg{[}-8e^{{z_1}/a(Q^2)}\textrm{Ei}(-z_{1}/a(Q^2))
\nonumber \\
&+&2e^{z_{2}/a(Q^2)}\textrm{Ei}(-z_{2}/a(Q^2))\nonumber \\
&+&16e^{-z_{1}/a(Q^2)}\textrm{Ei}(z_{1}/a(Q^2))
\nonumber \\
&-&10e^{-z_{2}/a(Q^2)}\textrm{Ei}(z_{2}/a(Q^2))\Bigg{]},
 \\
{\MU}^{(L)}_{PT}(Q^2)&=&\frac{1}{3b}\Bigg{[}8e^{-z_{1}/a(Q^2)}\textrm{Ei}(z_{1}/a(Q^2))
\nonumber \\
&-&6e^{-z_{2}/a(Q^2)}\textrm{Ei}(z_{2}/a(Q^2))
\nonumber \\
&-&2e^{z_{2}/a(Q^2)}\textrm{Ei}(-z_{2}/a(Q^2))\Bigg{]}.
\end{eqnarray}
\begin{figure*}[t]
\centering
\begin{tabular}{lr}
\psfrag{y}{\rotatebox{90}{$\mathcal{D}_{\tiny{\mbox{$PT$}}}^{\tiny{\mbox{($L$)}}}(Q^{2})$}}
\psfrag{x}{\mbox{$Q^{2}/\Lambda^{2}$}}
\includegraphics[width=0.45\textwidth]{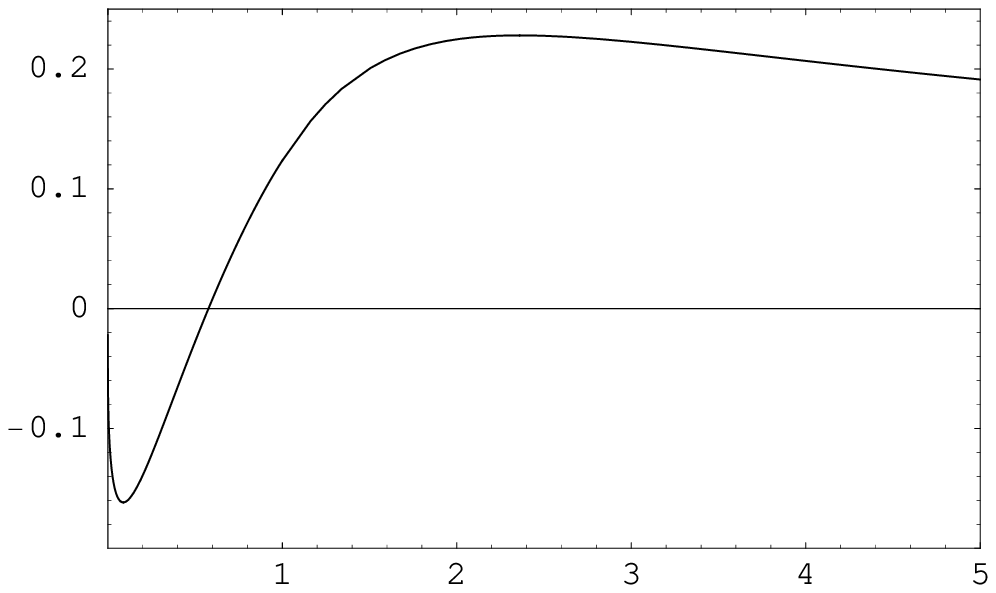}\hspace{.045\textwidth}&\hspace{.045\textwidth}
\psfrag{y}{\rotatebox{90}{$\mathcal{U}_{\tiny{\mbox{$PT$}}}^{\tiny{\mbox{($L$)}}}(Q^{2})$}}
\psfrag{x}{\mbox{$Q^{2}/\Lambda^{2}$}}
\includegraphics[width=0.45\textwidth]{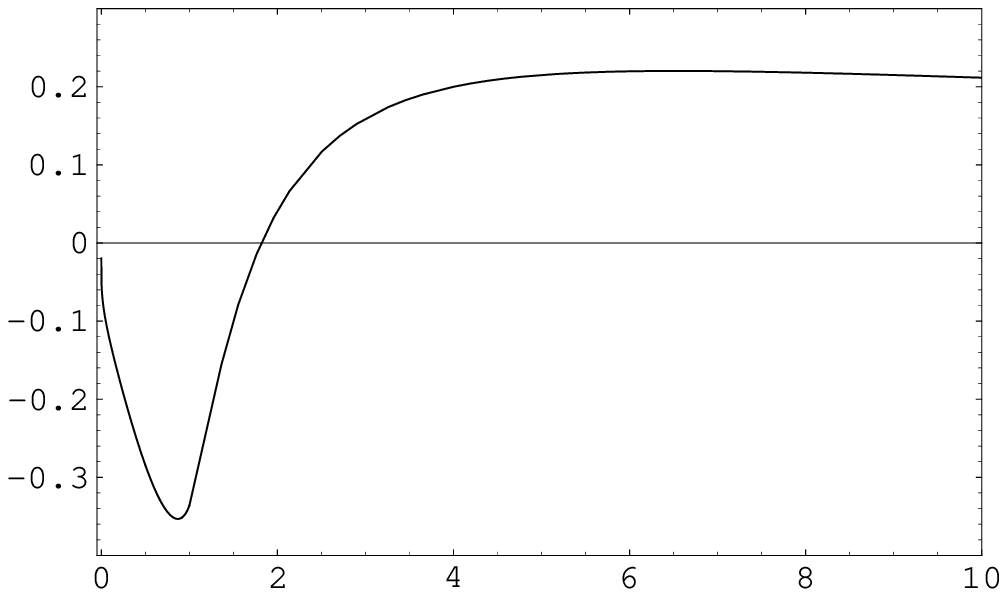}\\\\\\
\multicolumn{2}{c}{
\psfrag{y}{\rotatebox{90}{$\mathcal{K}_{\tiny{\mbox{$PT$}}}^{\tiny{\mbox{($L$)}}}(Q^{2})$}}
\psfrag{x}{\mbox{$Q^{2}/\Lambda^{2}$}}
\includegraphics[width=0.45\textwidth]{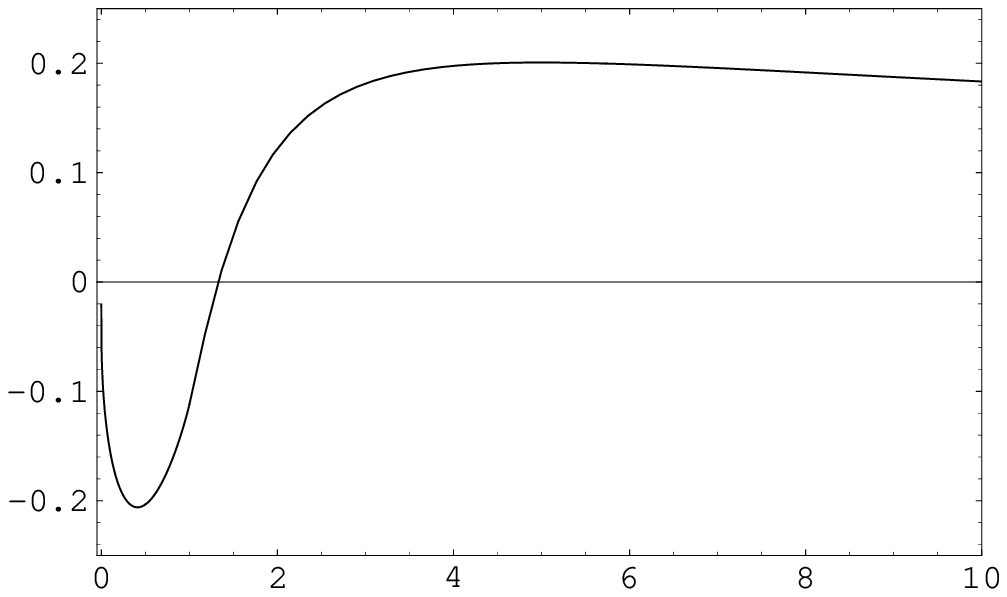}}
\end{tabular}
\caption{The perturbative corrections to the parton model result vanish in the vicinity of $Q^2=\Lambda^2$ and then freeze
smoothly to zero through negative values.}
\end{figure*}

These expressions have the property that they are finite and continuous at $Q^2=\Lambda^2$ where the one-loop coupling
$a(Q^2)=2/b\ln (Q^2/\Lambda^2)$ has a Landau pole. For $Q^2<\Lambda^2$ the standard Borel representation breaks down
and as we shall argue needs to be replaced by the modified Borel representation
\begin{eqnarray}
{\cal{D}}^{(L)}_{PT}({Q}^{2})=\int_{0}^{-\infty}{dz}\,{e}^{-z/a(Q^2)}B[{\cal{D}}^{(L)}_{PT}](z)\;.
\end{eqnarray}
We shall show that this pair of Borel representations are equivalent to the one-chain skeleton
expansion term in the two $Q^2$ regions. The $Q^2<\Lambda^2$ representation has an ambiguous imaginary part
due to the ${UV}_{n}$ singularities on the integration contour. 
In Fig.5 we show the $Q^2$ behaviour of the three Euclidean observables. The perturbative corrections
are seen to change sign in the vicinity of $Q^2={\Lambda}^{2}$, and there is then a smooth freezing behaviour
and they vanish as $Q^2\rightarrow 0$. 
The finiteness is delicate. For small $x$
\begin{eqnarray}
{\rm{Ei}}(x)=\ln |x|+{\gamma}_{E}+{\cal{O}}(x)
\end{eqnarray}
and so as $Q^2\rightarrow\Lambda^2$ there is potentially a $\ln a$ divergence. 
For $\MD$ the coefficient
of this divergent term is \cite{r1}
\begin{eqnarray}
-\sum_{n+1}^{\infty}{z}_{n}^{2}[A_1(n)+B_1(n)]
\end{eqnarray}
For ${\MK}^{(L)}_{PT}(Q^2)$ and ${\MU}^{(L)}_{PT}(Q^2)$ the equivalent coefficients are
$(-8+2=16-10=0)$ and $(8-6-2)=0$, respectively. There is a relation between IR and UV renormalon residues
which ensures the divergent term vanishes
\begin{eqnarray}
{z}^{2}_{n+3}B_1(n+3)=-{z}_{n}^{2}A_1(n).
\end{eqnarray}
This ensures that
\begin{eqnarray}
\sum_{n=1}^{\infty}{z}_{n}^{2}[A_1(n)+B_1(n)]=0.
\end{eqnarray}
Another similar relation is \cite{r14}
\begin{eqnarray}
{A}_{0}(n)=-{B}_{0}(n+2).
\end{eqnarray}
We shall show that these relations are underwritten by continuity of the characteristic
function in the skeleton expansion. At $Q^2=\Lambda^2$ one finds the finite values,
\begin{eqnarray}
{\MD}_{PT}^{(L)}(Q^{2}={\Lambda}^{2})&=&\sum_{n=1}^{\infty}z_{n}[A_{0}(n)-B_{0}(n)]
\nonumber \\
&-&\sum_{n=1}^{\infty}z^{2}_{n}[A_{1}(n)+B_{1}(n)]\ln
n\nonumber \\
&\approx&\frac{0.679938}{b}\;,
\end{eqnarray}
and
\begin{eqnarray}
\MK_{PT}^{(L)}(\QQ ={\Lambda}^{2})=-\frac{8}{9b}\ln2\;,\nonumber \\
\MU_{PT}^{(L)}(\QQ ={\Lambda}^{2})=-\frac{8}{3b}\ln2\;.
\end{eqnarray}
\subsection{QCD skeleton expansion}
In QED the insertion of chains of bubbles into a basic skeleton diagram produces
a well-defined skeleton expansion. In QCD by modifying the coefficient of $\ln(-k^2/Q^2)$
in $\Pi_0 (k^2)$ to involve the QCD beta-function coefficient $b$ one can recast Eq.(71)
in the form
\begin{eqnarray}
&&d_{n}^{(L)}a^{n+1}=
a\int_{0}^{\infty}{dk^2}
\omD\Bigg{(}\frac{k^{2}}{Q^{2}}\Bigg{)}\Big{(}-\frac{ba}{2}\ln{\Bigg{(}\frac{k^{2}}{Q^{2}}\Bigg{)}}
\Big{)}^{n},\nonumber \\
&&\Rightarrow
{\MD}^{(L)}_{PT}(Q^2)=
\nonumber \\
&&Q^{2}\int_{0}^{\infty}\frac{d^{2}k}{k^{2}}\omD\Bigg{(}\frac{k^{2}}{Q^{2}}\Bigg{)}\frac{k^{2}}{Q^{2}}
\Bigg{[}\frac{a}{1+\frac{ba}{2}\ln{\Big{(}\frac{k^{2}}{Q^{2}}}\Big{)}}\Bigg{]}, 
\end{eqnarray}
which, defining $t=k^2/Q^2$, can be written as 
\begin{eqnarray}
{\cal{D}}^{(L)}_{PT}(Q^2)=\int_{0}^{\infty}{dt}\;{\omD}(t)a({t}{Q}^{2})\;.
\end{eqnarray}
Here ${\omD}(t)$ is the {\it characteristic function}. It satisfies the normalization condition
\begin{eqnarray}
\int_{0}^{\infty}{dt}\;{\omD}(t)=1\;,
\end{eqnarray}
which ensures that the leading coefficient of unity in the perturbative expansion of Eq.(31)
is reproduced. 
$\omD{(t)}$ and its first three derivatives are piecewise continuous at $t=1$
and the function divides into an IR and a UV part
\begin{eqnarray}
{\MD}^{(L)}_{PT}=\int_{0}^{1}{dt}\omD^{IR}(t)a(tQ^2)+\int_1^{\infty}{dt}\omD^{UV}a(tQ^2)
\end{eqnarray}
We shall see that the IR ${k^2}<Q^2$) and UV (${k^2}>Q^2$) components respectively reproduce the IR renormalon and UV
renormalon contributions in the Borel plane. For $Q^2>\Lambda^2$ one encounters the Landau pole in the coupling
in the first IR integral at $t=\Lambda^2/Q^2$, and the integral requires regulation (e.g. PV).  This mirrors the IR renormalon
ambiguities in the Borel integral of Eq.(74). For $Q^2<\Lambda^2$ one encounters the Landau pole in the coupling
in the second UV integral, and this requires regulation. This mirrors UV renormalon ambiguities in the modified
Borel integral of Eq.(90).
$\omD (t)$ can be derived by using classic QED work of Baker and Johnson on vacuum polarization.
The vacuum polarization function of Eq.(22) can be written as
\begin{eqnarray}
\Pi(Q^{2})=\int_{0}^{\infty}{dt}\;\omP(t)a(t Q^{2})
\end{eqnarray}
where the characterisitic function $\omP(t)$ is given by
\begin{eqnarray}
\omP(t)=-\frac{4}{3}
\left\{
    \begin{array}{ll}
    t\,\Xi(t)                                    &\hskip10mm     t\leq1\quad\leftrightarrow\quad\textrm{IR} \\
                                                &  \label{eq:45}\label{conf}\\
    \frac{1}{t}\,\Xi\Big{(}\frac{1}{t}\Big{)}     &\hskip10mm     t\geq1\quad\leftrightarrow\quad\textrm{UV} \\
    \end{array}
\right.
\end{eqnarray}
${\Xi}(t)$ is given by
\begin{eqnarray}
\Xi(t)&\equiv&\frac{4}{3t}[1-\ln t+\Big{(}\frac{5}{2}-\frac{3}{2}\ln
t\Big{)}t
\nonumber \\
&+&\frac{(1+t)^{2}}{t}[L_{2}(-t)+\ln t\ln(1+t)]].
\end{eqnarray}
where ${L}_{2}(x)$ is the dilogarithmic function
\begin{eqnarray}
L_{2}(x)=-\int_{0}^{x}dy\frac{\ln(1-y)}{y}\;.
\nonumber
\end{eqnarray}
${\Xi}(t)$ corresponds to the Bethe-Salpeter kernel for the scattering of light-by-light
involving the diagrams in Fig.6. 
\begin{figure*}[t]
\centering
\begin{tabular}{lcr}
\includegraphics[width=0.3\textwidth]{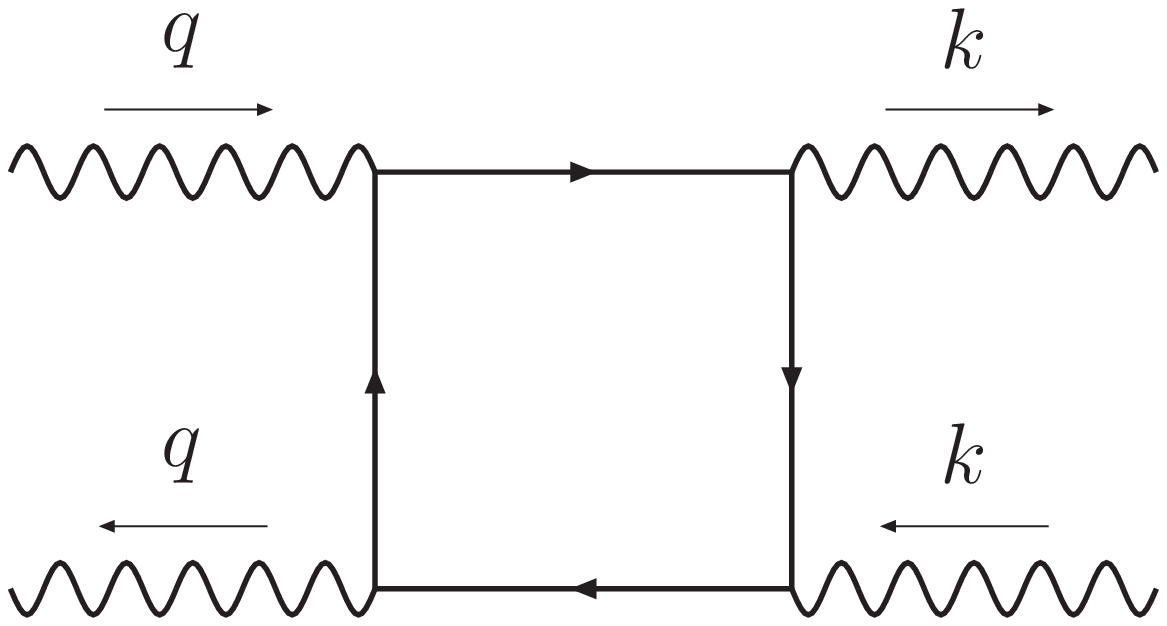}&\includegraphics[width=0.3\textwidth]{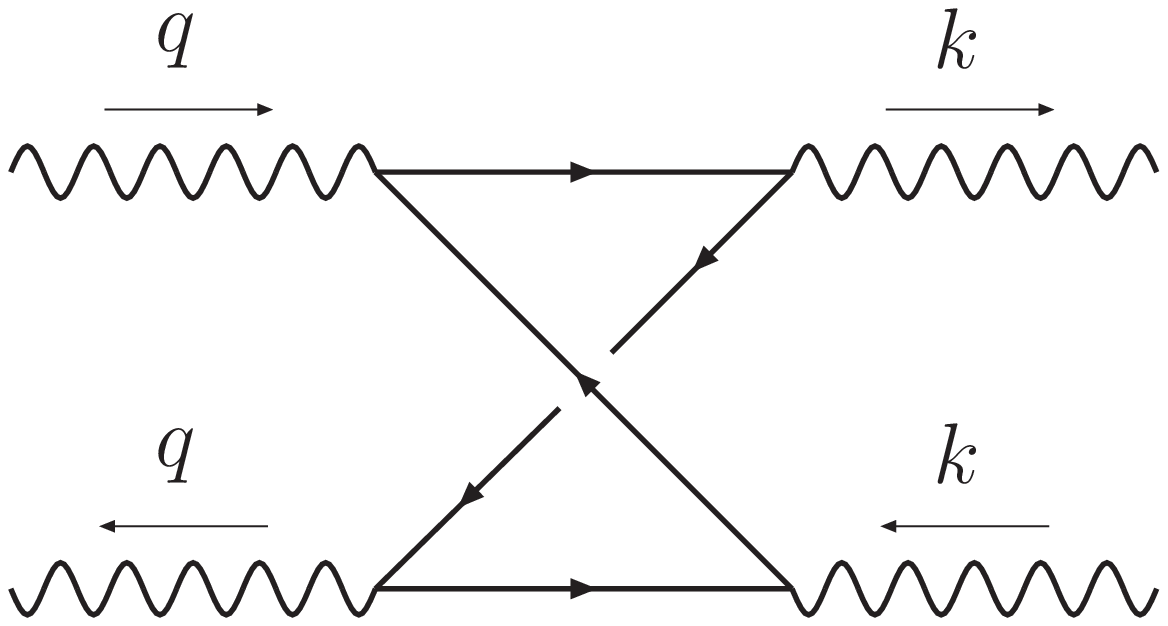}&{\includegraphics[width=0.3\textwidth]{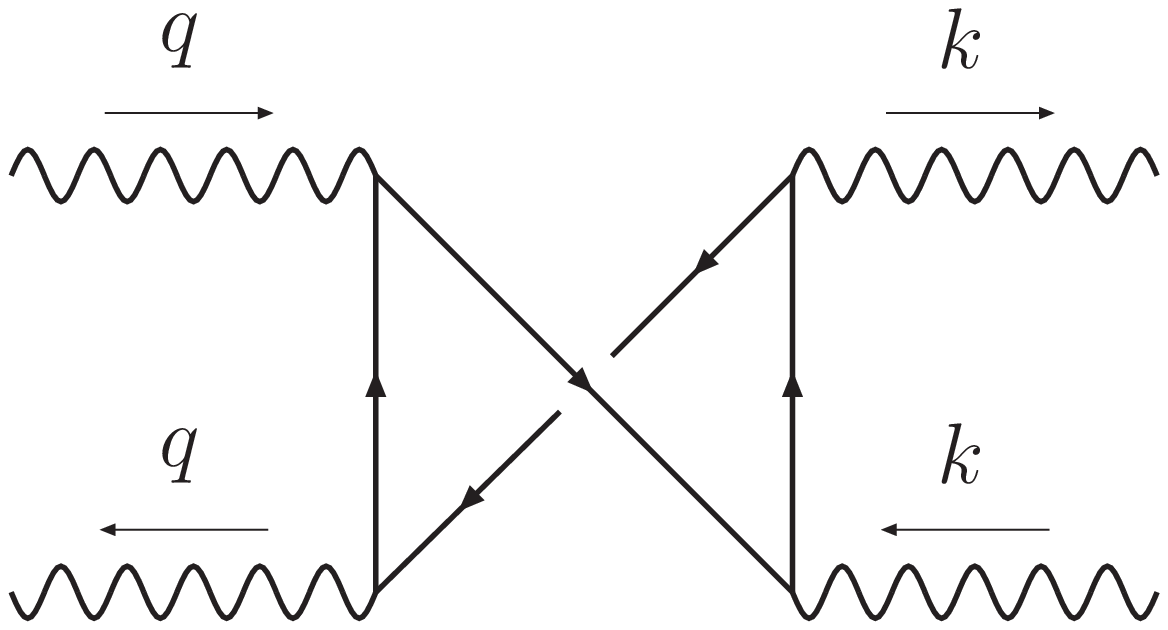}}\\
\end{tabular}
\caption{Light-by-light scattering diagrams, used to calculate $\omP$.}
\label{F:Kernel}
\end{figure*}
Notice that by attaching the ends of the fermion bubble chain to the momentum $k$
external propagators in Fig.6 one reproduces the topology of the diagrams in Fig.2.
This computation of $\omP$ is a {\it one-loop} calculation, and as we shall see
can be converted directly into the Borel plane renormalon structure for the $D$-function. This provides
a much simpler route than the full two-loop calculation of Refs.\cite{r11,r12}.

Changing from $\Pi$ to $\MD$ induces a transformation in ${\omP}$ of
\begin{eqnarray}
\PIQ&\rightarrow&Q^{2}\frac{d}{dQ^{2}}\PIQ=-\frac{4}{3}\MD(Q^{2})\nonumber \\
\Rightarrow\qquad\omP(t)&\rightarrow&\omP(t)+t\frac{d}{dt}\omP(t)=-\frac{4}{3}\omD(t).
\end{eqnarray}
One can write $\omP (t)$ as an expansion in powers of $t$ and $\ln t$ times powers of $t$
\begin{eqnarray}
\omP^{IR}(t)&=&-\frac{4}{3}\left(\sum_{n=1}^{\infty}\xi_{n}t^{n}+\ln t\sum_{n=2}^{\infty}\hat{\xi}_{n}t^{n}\right)\;,
\end{eqnarray}
and the conformal symmetry $t\leftrightarrow\frac{1}{t}$ between the UV and IR regions
means that the UV part can be written in terms of the same coefficients
\begin{eqnarray}
\omP^{UV}(t)&=&-\frac{4}{3}\left(\sum_{n=1}^{\infty}\xi_{n}t^{-n}-\ln t\sum_{n=2}^{\infty}\hat{\xi}_{n}t^{-n}\right)\;.
\end{eqnarray}
The $\xi_n$ and ${\hat{\xi}}_{n}$ are found to be
\begin{eqnarray}
\xi_{n>1}&=&\frac{4}{3}\frac{(2-6n^{2})(-1)^{n}}{(n-1)^{2}n^{2}(n+1)^{2}},
\nonumber \\
\hat{\xi}_{n>1}&=&\frac{4}{3}\frac{2(-1)^{n}}{(n-1)n(n+1)}\label{xi}\nonumber \\
\xi_{1}&=&1\;\hat{\xi}_{1}=0.
\end{eqnarray}
For $\omD$ we have
\begin{eqnarray}
\omD^{IR}(t)&=&\sum_{n=1}^{\infty}[\xi_{n}(1+n)+\hat{\xi}_{n}]t^{n}
\nonumber \\
&+&\ln t\sum_{n=2}^{\infty}\hat{\xi}_{n}(n+1)t^{n}\nonumber \\
\omD^{UV}(t)&=&\sum_{n=1}^{\infty}[\xi_{n}(1-n)-\hat{\xi}_{n}]t^{-n}
\nonumber \\
&+&\ln
t\sum_{n=2}^{\infty}\hat{\xi}_{n}(n-1)t^{-n}.
\end{eqnarray}
By making the change of variables $z=-a(Q^2)(n+1)\ln t$ for $t<1$ and $z=a(Q^2)(n-1)\ln t$ for
$t>1$, one can transform the skeleton expansion result into the standard Borel representation
of Eqs.(74,75).  
For $Q^2<\Lambda^2$, $a(Q^2)<0$, and one obtains the modified Borel representation of Eq.(90)
in which the upper limit of integration in $z$ is $-\infty$. For consistency one requires relations
between the residues $A_{0,1}$, $B_{0,1}$, and the characteristic function coefficients ${\xi}_{n}$, ${\hat{\xi}}_{n}$.
\begin{eqnarray}
\frac{\xi_{n}(1+n)+\hat{\xi}_{n}}{n+1}&=&-B_{1}(n+1)z_{n+1}\;\;\;  n\geq1\nonumber \\
\nonumber \\
\frac{\xi_{n}(1-n)-\hat{\xi}_{n}}{n-1}&=&A_{1}(n-1)z_{n-1}\,\,\,
n\geq2 
\end{eqnarray}
for the single pole residues and
\begin{eqnarray}
-\frac{\hat{\xi}_{n}(n+1)}{(n+1)^{2}}&=&B_{0}(n+1)+B_{1}(n+1)z_{n+1}\;\;\;
n\geq2
\nonumber \\
\frac{\hat{\xi}_{n}(n-1)}{(n-1)^{2}}&=&A_{0}(n-1)-A_{1}(n-1)z_{n-1}\;\;\;
n\geq2
\end{eqnarray}
for the double pole residues. 
These relations may be used to rewrite the series for $\omD(t)$ in terms of the residues
${A}_{0,1},{B}_{0,1}$, 
\begin{eqnarray}
\!\!\omD^{IR}(t)\!\!&=&\!\!\frac{b}{2}\sum_{n=1}^{\infty}-{z}_{n+1}^{2}{B}_{1}(n+1){t}^{n}
\nonumber \\
&-&{\ln t}\sum_{n=2}^{\infty}
{(n+1)}^{2}[{B}_{0}(n+1)
\nonumber \\
&+&{z}_{n+1}{B}_{1}(n+1)]{t}^{n}\;\; \nonumber \\
\!\!\omD^{UV}(t)\!\!&=&\!\!\frac{b}{2}\sum_{n=1}^{\infty}{z}_{n-1}^{2}{A}_{1}(n-1){t}^{-n}
\nonumber \\
&+&{\ln t}\sum_{n=2}^{\infty}{(n-1)}^{2}[{A}_{0}(n-1)
\nonumber \\
&-&{z}_{n-1}{A}_{1}(n-1)]{t}^{-n}.\;\;
\end{eqnarray}
One can then show that continuity of $\omD (t)$ and its first three derivatives at
$t=1$, and equivalently finiteness of ${\MD}^{(L)}_{PT}(Q^2)$ and its first three derivatives $d/d\ln Q$
at $Q^2=\Lambda^2$ is underwritten by the following relations between the ${A}_{0,1}$ and ${B}_{0,1}$ residues
\begin{eqnarray}
\sum_{n=1}^{\infty}{z}_{n}^{2}[{A}_{1}(n)+B_{1}(n)]=0.
\end{eqnarray} 
This is just Eq.(94) which guarantees finiteness of ${\MD}(Q^2)$ at $Q^2=\Lambda^2$. In addition we
have more complicated relations which underwrite continuity and finiteness of the derivatives.\cite{r1}
\begin{eqnarray}
&&\sum_{n=1}^{\infty}[2{z}_{n}^{3}({A}_{1}(n)
\nonumber \\
&-&{B}_{1}(n))-{z}_{n}^{2}({A}_{0}(n)+{B}_{0}(n)]=0\;.
\\
&&\sum_{n=1}^{\infty}[3{z}_{n}^{4}({A}_{1}(n)+{B}_{1}(n))
\nonumber \\
&-&2{z}_{n}^{3}({A}_{0}(n)-{B}_{0}(n))]=0\;.
\\
&&\sum_{n=1}^{\infty}[4{z}_{n}^{5}(A_1(n)-B_1(n))
\nonumber \\
&-&3{z}_{n}^{4}(A_0(n)+B_0(n))]=0\;.
\end{eqnarray}
\begin{figure*}[t]
\centering
\begin{tabular}{lr}
\psfrag{y}{\rotatebox{90}{$\MD^{\tiny{\mbox{($L$)}}}(Q^{2})$}}
\psfrag{x}{\mbox{$Q^{2}/\Lambda^{2}$}}
\includegraphics[width=0.45\textwidth]{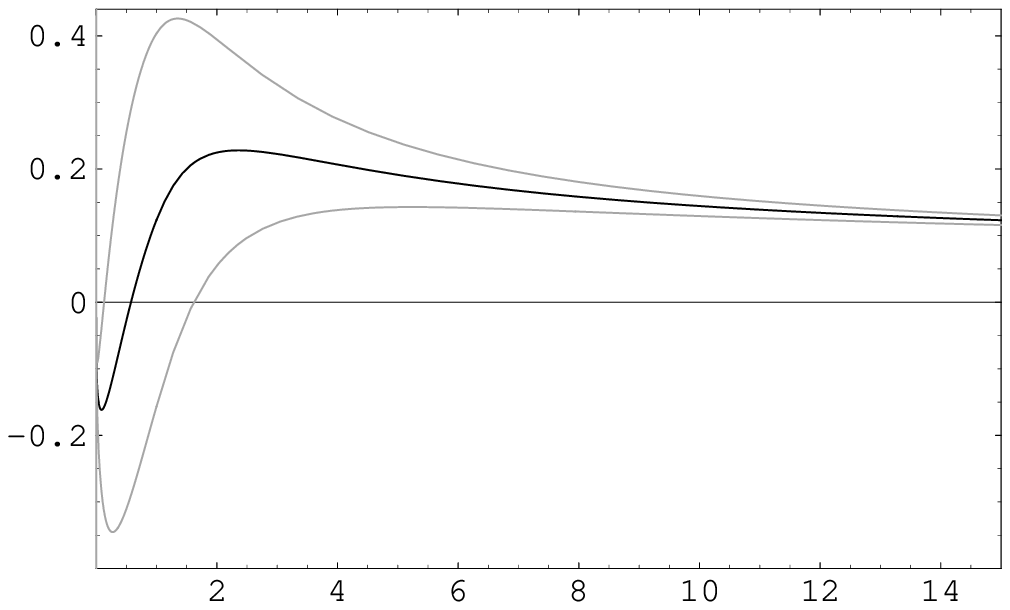}\hspace{.045\textwidth}&\hspace{.045\textwidth}
\psfrag{y}{\rotatebox{90}{$\MU^{\tiny{\mbox{($L$)}}}(Q^{2})$}}
\psfrag{x}{\mbox{$Q^{2}/\Lambda^{2}$}}
\includegraphics[width=0.45\textwidth]{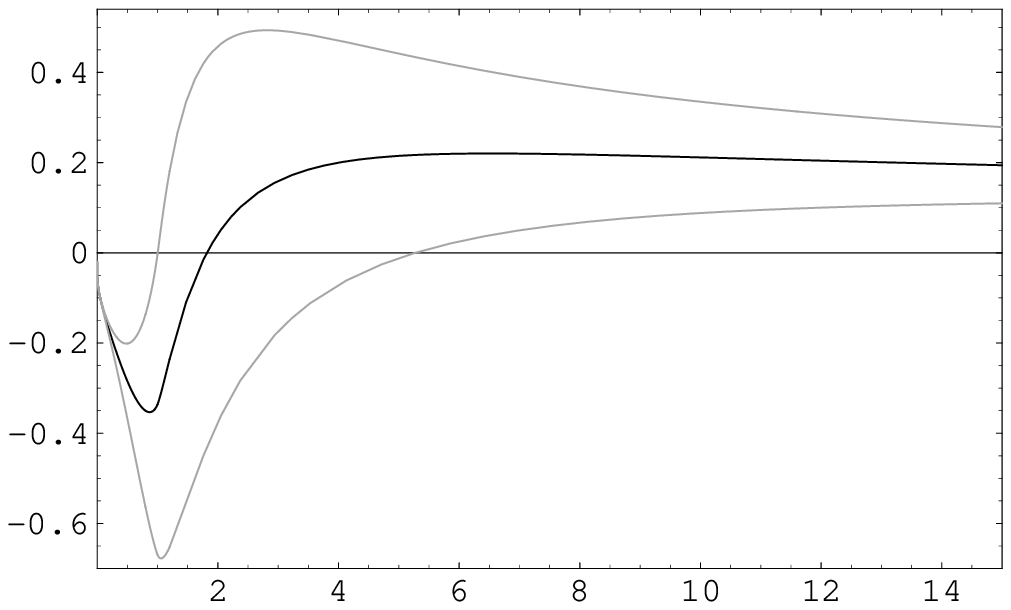}\\\\\\
\multicolumn{2}{c}{\psfrag{y}{\rotatebox{90}{$\MK^{\tiny{\mbox{($L$)}}}(Q^{2})$}}
\psfrag{x}{\mbox{$Q^{2}/\Lambda^{2}$}}
\includegraphics[width=0.45\textwidth]{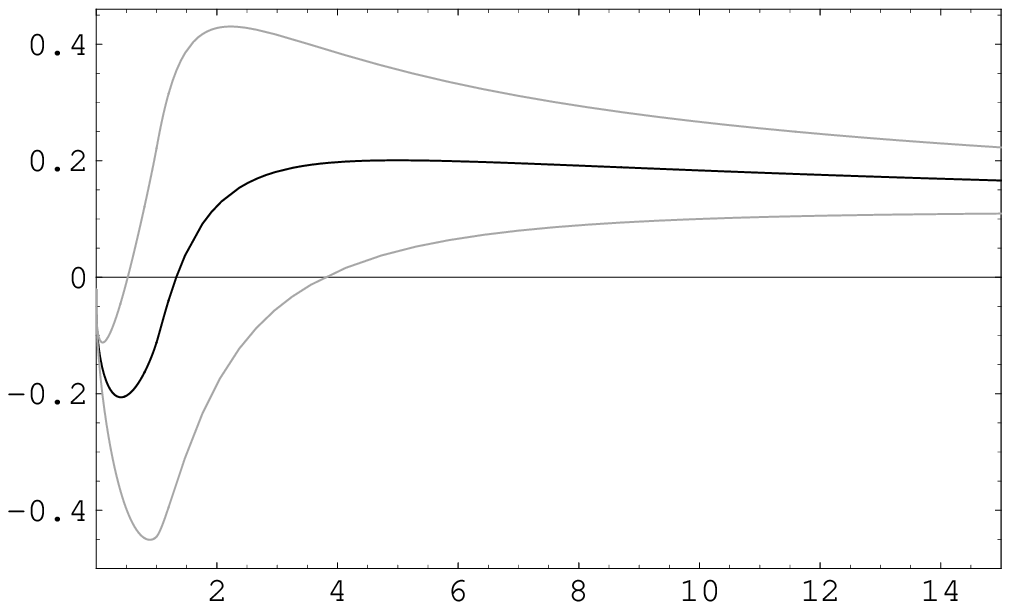}}
\end{tabular}
\caption{The bold curves show the choice $\kappa=0$, i.e. just the PT component as in the earlier
plots. The upper and lower curves correspond to the choices $\kappa=1$ and $\kappa=-1$, respectively.} 
\end{figure*}
\subsection{The NP component}
It is easy to show \cite{r1} that the ambiguous imaginary part in ${\MD}^{(L)}_{PT}$ arising from IR renormalons
for $Q^2>\Lambda^2$ and UV renormalons for $Q^2<\Lambda^2$ can be written directly in terms of $\omD^{IR}$
and $\omD^{UV}$,
\begin{eqnarray}
Im[{\MD}^{(L)}_{PT}(Q^2)]&=&\pm \frac{2\pi}{b}\frac{{\Lambda}^{2}}{Q^2}\omD^{IR}\left(\frac{\Lambda^{2}}{Q^2}\right)\;\;\;(Q^2>{\Lambda}^{2}) \nonumber \\
Im[{\MD}^{(L)}_{PT}(Q^2)]&=&\pm\frac{2\pi}{b}\frac{{\Lambda}^{2}}{Q^2}\omD^{UV}\left(\frac{\Lambda^{2}}{Q^2}\right)\;\;\;(Q^2<{\Lambda}^{2})
\end{eqnarray}
Continuity at $Q^2=\Lambda^2$ then follows from continuity of $\omD (t)$ at $t=1$. In principle the real part
of the OPE condensates are independent of the imaginary, but continuity and finiteness involve the set
of relations between ${A}_{0,1}$ and ${B}_{0,1}$ in Eqs.(113-116). Although not strictly necessary for continuity, 
continuity naturally follows if we write 
\begin{eqnarray}
{\MD}^{(L)}_{NP}(Q^2)=\left(\kappa\pm\frac{2\pi i}{b}\right)\int_{0}^{{\Lambda}^2/Q^2}{dt}\;\left({\omD}(t)+t\frac{d\omD(t)}{dt}\right)\;.
\end{eqnarray}
Here $\kappa$ is an overall real non-perturbative constant. If the PT component is PV regulated then one
averages over the $\pm$ possibilities for contour routing, combining with ${\MD}^{(L)}_{PT}$ one can then write down
the overall result for ${\MD}^{(L)}(Q^2)$.
\begin{eqnarray}
{\MD}^{(L)}(Q^2)&=&\int_{0}^{\infty}{dt}\;\Bigg[{\omD}(t)a(tQ^2)
\nonumber \\
&+&\kappa\left(\omD(t) + t\frac{d\omD (t)}{dt}\right)\theta({\Lambda}^{2}-tQ^2)\Bigg]\;. 
\end{eqnarray}
The $Q^2$ evolution is fixed by the non-perturbative constant $\kappa$ and by $\Lambda$.
The evolution for the choices $\kappa=0,1,-1$ is shown in Fig.7. 
We see that the QCD corrections to the parton model result
corrections freeze smoothly to zero as $Q^2\rightarrow 0$.
\section{Concluding remarks}
In these lectures we have shown how the large-order growth of perturbative coefficients
in QED and QCD is dominated by renormalon contributions associated with diagrams such
as those in Figs.2,4, in which chains of fermion bubbles are inserted inside a basic
skeleton diagram. We have focused on the vacuum polarization Adler function $D(Q^2)$
as our main example, with a cursory consideration of DIS sum rules as well. In QCD one
is forced to use the so-called leading-$b$ approximation in which one inserts chains
of effective bubbles in which the logarithm in the renormalised $\Pi_0(k^2)$ has a
coefficient of $b$, since corrections to the gluon propagator are gauge-dependent.
In the approximation in which a single chain is inserted one finds single and double
pole IR and UV renormalon singularities in the Borel plane for Adler-$D$ and for the
DIS sum rules. Interestingly, these one-chain approximation results are finite
and continuous at $Q^2=\Lambda^2$ where the coupling has a Landau pole, and this
property is underwritten by relations between IR ($k^2<Q^2$) and UV ($k^2>Q^2$) physics,
where $k^2$ refers to the momentum flowing through the bubble chain. There are corresponding
relations between IR and UV renormalon residues. These connections are more transparent
in the language of the one-chain skeleton expansion of Eq.(99) in which they are seen to
correspond to continuity of the characteristic function $\omega (t)$ and its derivatives.
Perturbative and non-perturbative ambiguities which contribute an imaginary part may also
be written in terms of the characteristic function.  
Presumably the one-chain approximation
is far too crude to describe the real non-perturbative infra-red behaviour of QCD observables
, but is interesting that it does have the properties of continuity and finiteness that
must be possessed by the true all-orders result. It will be interesting in future studies
to see how these properties arise if one includes higher numbers of chains, and also
instanton effects. 

\begin{acknowledgements}
Yasaman Farzan and the rest of the organising committee of the IPM LPH-06 meeting 
are thanked for their
painstaking organisation of this stimulating and productive school and conference.
Many thanks are also due to Abolfazl Mirjalili for organising my wonderful post-conference
visits to Esfahan, Yazd and Shiraz, and to all those whose welcoming hospitality
made my first visit to Iran so extremely enjoyable.
\end{acknowledgements}

\end{document}